\documentstyle{lamuphys}
\begin{document}
%
%%%%%%%%%%%%%%%%%%%%%%%%%
% uncomment the next 14 lines to get page numbering right when
% printing a two-up version:
%\thispagestyle{empty}
%\noindent
%\rule{\textwidth}{1pt}
%\vspace{2pt}
%
%\noindent
%{\Huge Exact S-matrices}
%
%\vspace{5pt}
%\noindent
%\rule{\textwidth}{1pt}
%\clearpage
%
%%%%%%%%%%%%%%%%%%%%%%%%%
%% my macros
\newcommand{\nn}{\nonumber}
\newcommand{\bea}{\begin{eqnarray}}
\newcommand{\eea}{\end{eqnarray}}
\newcommand{\half}{\frac{1}{2}}
\newcommand{\fract}[2]{{\textstyle\frac{#1}{#2}}}
\newcommand{\bra}[1]{\langle{#1}|}
\newcommand{\ket}[1]{|{#1}\rangle}
\newcommand{\exercise}[1]{\noindent%
{\small\trivlist\item[]Exercise:#1\endtrivlist}}
\newcommand{\bexercise}[1]{\noindent%
{\small\trivlist\item[](Exercise:#1)\endtrivlist}}
\newcommand{\sP}{s^+}
\newcommand{\sM}{s^-}
\newcommand{\im}{{\rm Im}\,}
\newcommand{\NP}[1]{Nucl.\ Phys.\ {\bf #1}}
\newcommand{\HPA}[1]{Helv.\ Phys.\ Acta\ {\bf #1}}
\newcommand{\AP}[1]{Ann.\ Phys.\ {\bf #1}}
\newcommand{\PL}[1]{Phys.\ Lett.\ {\bf #1}}
\newcommand{\NC}[1]{Nuovo Cim.\ {\bf #1}}
\newcommand{\CMP}[1]{Comm.\ Math.\ Phys.\ {\bf #1}}
\newcommand{\PR}[1]{Phys.\ Rev.\ {\bf #1}}
\newcommand{\PRL}[1]{Phys.\ Rev.\ Lett.\ {\bf #1}}
\newcommand{\PRP}[1]{Phys.\ Rep.\ {\bf #1}}
\newcommand{\PTP}[1]{Prog.\ Theor.\ Phys.\ {\bf #1}}
\newcommand{\PTPS}[1]{Prog.\ Theor.\ Phys.\ Supp.\ {\bf #1}}
\newcommand{\MPL}[1]{Mod.\ Phys.\ Lett.\ {\bf #1}}
\newcommand{\IJMP}[1]{Int.\ J.\ Mod.\ Phys.\ {\bf #1}}
\newcommand{\ETDS}[1]{Ergod.\ Th.\ \& Dynam.\ Sys.\ {\bf #1}}
\newcommand{\JETP}[1]{Sov.\ Phys.\ JETP {\bf #1}}
\newcommand{\JETPL}[1]{JETP Lett.\ {\bf #1}}
\newcommand{\TMP}[1]{Teor.\ Math.\ Phys.\ {\bf #1}}
\newcommand{\TMF}[1]{Teor.\ Mat.\ Fiz.\ {\bf #1}}
\hyphenation{Zamo-lod-chikov}
%
%%%% macros for root tables
%
\newcommand{\B}{\bullet}
\newcommand{\cd}{$\cdot~$}
\newcommand{\gap}{\cd\cd\cd\cd\cd}
\newcommand{\dfil}{\leaders\hbox to .1in{\hfil$\cdot\,$}\hfil\hskip -.005in}
\newcommand{\hs}[1]{\hbox to .5in{\dfil$#1$}}
\newcommand{\rs}[1]{\hbox to .5in{\hfil$\a_{#1}$}}
\newcommand{\Label}[1]{\llap{$\scriptstyle #1$}}
\newcommand{\N}{\hbox to .5in{\dfil}}
\newcommand{\Y}{\hs{\B}}
\newcommand{\YY}{\hs{\B\B}}
\newcount\n
\newcount\m
\newcommand{\y}[1]{\hs{\m=0\loop\ifnum\m<#1\advance\m by1{\mkern
-1.5mu\B}\repeat}}
\newcommand{\ys}[2]{\n=0\loop\ifnum\n<#2\advance\n by1\y#1\repeat}
\newcommand{\Ys}[1]{\n=0\loop\ifnum\n<#1\advance\n by1\Y\repeat}
\newcommand{\Ns}[1]{\n=0\loop\ifnum\n<#1\advance\n by1\N\repeat}
\newcommand{\As}[1]{\n=0\loop\ifnum\n<#1\advance\n by1\rs{\the\n}\repeat}
\newcommand{\row}[2]{\hbox{\Label{#1~}#2}}
%
%%%% line with ticks
%
\newcommand{\tick}[2]{\vrule height#1pt depth#2pt}
\newcommand{\seg}[2]{{\hfil \tick#1.#2.}}
\newcount\n
\newcommand{\tline}[3]{{\n=0 \loop\ifnum\n<#1 \advance\n by1 {\hfil
   \tick{#2}{#3}}\repeat}}
\newcommand{\tickline}[6]
 {\n=0 \vbox{\hrule \vskip-#3pt
   \hbox to #1in{\label{#5}\tick{#3}{#4}
   \loop\ifnum\n<#2 \advance\n by1 \seg{#3}{#4}\repeat\Label{#6}}}}
%
%%%% macros for brick walls
%
\newdimen\bw
\newdimen\hbw
\newdimen\tw
\bw=9.0pt
\hbw=4.5pt
\newcommand{\ha}{\hskip .5pt}
\tw=150pt
\newcommand{\bx}{\vbox{\hrule\hbox to \bw{\vrule height 3.0pt%
\hfil\vrule}}}
\newcommand{\bxs}[1]{\n=0\loop\ifnum\n<#1\advance\n by1\bx\repeat}
\newcommand{\nbx}{\ha\hskip\bw\ha}
\newcommand{\p}[1]{\ha\vbox{\offinterlineskip\bxs{#1}}\ha}
\newcommand{\q}[1]{\n=0\loop\ifnum\n<#1\advance\n by1\nbx\repeat}
\renewcommand{\P}{\p1}
\newcommand{\nt}{\tline{9}{0}{1}\hfil}
\newcommand{\bag}[1]{\vbox{\offinterlineskip
\hbox{#1}
\hrule
\hbox to \tw{\tick{0}{2}\nt\tick{0}{2}\nt\tick{0}{2}\nt\tick{0}{2}}
}}
\newcommand{\bagh}[1]{\bag{\hskip\hbw\ha#1\ha\hskip\hbw}}
%
%%%%%%%%%%%%%%%%%%%%%%%%%
%
% START OF PAPER
%
%%%%%%%%%%%%%%%%%%%%%%%%%

\pagenumbering{arabic}
\setcounter{page}{1}

\title{Exact S-matrices}

\author{Patrick\,Dorey\inst{}}

\institute{
Department of Mathematical Sciences,\\
University of Durham, Durham DH1 3LE,\\
England\\
{\tt p.e.dorey@durham.ac.uk}}

\maketitle

\begin{abstract}
The aim of these notes is to provide an elementary introduction to some of
the basic elements of exact S-matrix theory. This is a large
subject, and only the beginnings will be covered here. A particular
omission is any serious
discussion of the Yang-Baxter equation; instead, the
focus will be on questions of analytic structure, and the
bootstrap equations.
Even then, what I have to say will only be a sketch of the simpler
aspects. The hope is to give a hint
of the many curious features of scattering theories in 1+1 dimensions.

\hfill DTP-98/69; {\tt hep-th/9810026}
 
% Based on lectures given in Budapest, August 1996 and also in Paris,
% November/December 1996.

\end{abstract}
\section{Introduction -- what's so special about 1+1?  }
To get things started, I want to describe a particularly simple
calculation that can be done in probably the simplest nontrivial
quantum field theory imaginable, 
namely $\lambda\phi^4$ theory in a universe with only one
spatial dimension.
 
The Lagrangian to consider is
\[
{\cal L}=\frac{1}{2}(\partial\phi)^2-\frac{1}{2}
m^2\phi^2-\frac{\lambda}{4!}\phi^4~,
\]
resulting in the Feynman rules

\begin{picture}(300,70)(0,-70)
\thinlines
\put(100,-20){\line(1,0){30}}
\put(140,-22){$\displaystyle{}=~\frac{i}{p^2-m^2+i\epsilon}$}
\put(105,-60){\line(1,1){20}}
\put(125,-60){\line(-1,1){20}}
\put(115,-50){\circle*{3}}
\put(140,-52){$\displaystyle{}=~{-i}\lambda$}
\end{picture}

The task is to calculate the connected $2\rightarrow 4$ production
amplitude, at tree level. Actually, to keep track of the diagrams it
is a little easier to look at the $3\rightarrow 3$ process, leaving
implicit the understanding that one of the {\it out} momenta will be
crossed to {\it in} at the end. I'll label the three {\it in}
particles as $a$, $b$, $c$, and the three {\it out} particles as
$d$, $e$, $f$, and opt to cross $c$ from {\it in} to {\it out} later.
It also helps to adopt light-cone
coordinates from the outset, using 
\[
(p,\bar p)=(p^0{+}p^1,p^0{-}p^1)
\]
and then solving the mass-shell condition $p\bar p=m^2$ by writing
the {\it in} and {\it out} momenta as
\[
p_a = (ma,ma^{-1})\quad,\quad
p_b = (mb,mb^{-1})
\]
and so on, with $a,b,\dots$ real numbers, positive for particles
travelling forwards in time. In terms of these variables, the crossing 
from $3\rightarrow 3$ to $2\rightarrow 4$ amounts to a continuation
from $c$ to $-c$.
For the $3\rightarrow 3$ amplitude there are just two
classes of diagram:

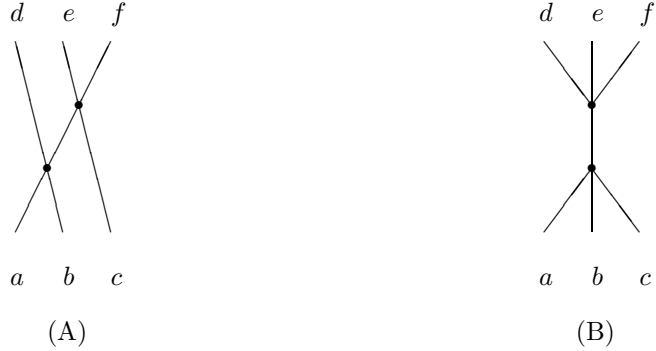
\begin{picture}(300,146)(10,572)
\thinlines
\put( 72,644){\circle*{3}}
\put( 84,668){\circle*{3}}
\put(278,644){\circle*{3}}
\put(278,668){\circle*{3}}
\put( 58,600){$a$}
\put( 78,600){$b$}
\put( 96,600){$c$}
\put(258,600){$a$}
\put(278,600){$b$}
\put(296,600){$c$}
\put( 58,700){$d$}
\put( 78,700){$e$}
\put( 96,700){$f$}
\put(258,700){$d$}
\put(278,700){$e$}
\put(296,700){$f$}
\put( 72,579){(A)}
\put(272,579){(B)}
\put( 60,620){\line( 1, 2){ 36}}
\put( 96,620){\line(-1, 4){ 18}}
\put( 78,620){\line(-1, 4){ 18}}
\put(260,620){\line( 3, 4){ 18}}
\put(278,644){\line( 3,-4){ 18}}
\put(278,620){\line( 0, 1){ 72}}
\put(260,692){\line( 3,-4){ 18}}
\put(278,668){\line( 3, 4){ 18}}
\end{picture}

\noindent
The internal momentum in (A) is
$p=m(a{+}b{-}d,a^{-1}{+}b^{-1}{-}d^{-1})$, and so its propagator
contributes
\begin{eqnarray}
\frac{i}{p^2-m^2}&=&\frac{i}{m^2}\frac{1}{(a{+}b{-}d)
(a^{-1}{+}b^{-1}{-}d^{-1}) - 1} \nonumber \\
&=&\frac{i}{m^2}\frac{-abd\phantom{-}}{(a{+}b)(a{-}d)(b{-}d)} 
\nonumber
\end{eqnarray}
to the total scattering amplitude. Given the agreement
above that one of the {\it out} momenta is actually in-going, this
propagator is never on-shell, and so forgetting about the
$i\epsilon$ does not cause any error. The same remark applies to
diagram (B), for which
\begin{eqnarray}
\frac{i}{p^2-m^2}&=&\frac{i}{m^2}\frac{1}{(a{+}b{+}c)
(a^{-1}{+}b^{-1}{+}c^{-1}) - 1} \nonumber \\
&=&\frac{i}{m^2}\frac{abc}{(a{+}b)(a{+}c)(b{+}c)}~. \nonumber
\end{eqnarray}
Adding these together, with a brief pause to check that the
diagrams have been counted correctly, yields the full result at tree
level:
\[
\langle{out}|{in}\rangle_{\rm tree}=
-\frac{i\lambda^2}{m^2}A_{\rm legs}H(a,b,c,d,e,f)
\]
where $A_{\rm legs}$ contains all the factors living on external legs
and so on that will be the same for all diagrams,
and
\[
H(a,b,c,d,e,f)=\sum_{{cycl \{abc\}}\atop{cycl\{def\}}}
\frac{-abd\phantom{-}}{(a+b)(a-d)(b-d)}+\frac{abc}{(a+b)(b+c)(c+a)}~,
\]
with the sum running over all cyclic permutations of $\{a,b,c\}$ and
$\{d,e,f\}$.

Now I need the following fact:
\[
\hbox{If}\qquad
a+b+c=d+e+f\quad\hbox{and}\quad 
\frac{1}{a}+\frac{1}{b}+\frac{1}{c}=
\frac{1}{d}+\frac{1}{e}+\frac{1}{f}
\]
\[
\hbox{then}\qquad
H(a,b,c,d,e,f)\equiv -1~.
\]
The two conditions are the so-far ignored conservation
of left- and right- lightcone momenta. The formula
makes no mention of the signs of the arguments to $H$, and certainly
holds with $c$ negative. The conclusion:

\smallskip
\noindent{$\bullet$} In 1+1-dimensional $\lambda\phi^4$ theory, the
$2\rightarrow 4$ amplitude is a {\it constant} at tree level.
\smallskip
 
\noindent
It is now very tempting to cancel this amplitude completely, by
adding a term
\[
-\frac{1}{6!}\frac{\lambda^2}{m^2}\phi^6
\]
to the original Lagrangian. In 1+1 dimensions this does not spoil
renormalisability, and gives a theory in which the $2\rightarrow 4$
amplitude {\it vanishes} at tree level. With $\beta^2=\lambda/m^2$,
the new Lagrangian is
\[
{\cal L}=\frac{1}{2}(\partial\phi)^2-
\frac{m^2}{\beta^2}\left[
\frac{1}{2}
\beta^2\phi^2+\frac{1}{4!}\beta^4\phi^4+
\frac{1}{6!}\beta^6\phi^6\right]~.
\]

This is already curious, but it is possible to go much further. Calculating
the $2\rightarrow 6$ amplitude (left as an exercise for the energetic
reader) should
reveal that {\it this} is now constant, ready to be killed off by a
judiciously-chosen $\phi^8$ term, and so on. At each stage a
residual constant piece can be removed by a (uniquely-determined)
higher-order interaction.
Keep going, and infinitely-many diagrams later you should find
\[
{\cal L}=\frac{1}{2}(\partial\phi)^2-
\frac{m^2}{\beta^2}\left[\cosh(\beta\phi)-1\right]~,
\]
the sinh-Gordon Lagrangian. Sending $\beta$ to $i\beta$
converts this into the well-known sine-Gordon model, to which the
discussion will return in later lectures.

The claim of uniqueness just made deserves a small 
caveat. I began the calculation with no $\phi^3$ term in the initial
Lagrangian, and a discrete $\phi\rightarrow-\phi$ symmetry which
persisted throughout. But what if I had instead started with a nonzero
$\phi^3$ term,
and tried to play the same game? This is definitely a harder problem,
but the final answer can be predicted with a fair degree of
confidence:
\bea
{\cal L}
&=&\frac{1}{2}(\partial\phi)^2-\half m^2\phi^2-
\frac{1}{3!}\lambda\phi^3-
\frac{1}{4!}\frac{3\lambda^2}{m^2}\phi^4-\dots\nn\\
&=&\frac{1}{2}(\partial\phi)^2-
\frac{m^2}{6\beta^2}\left[e^{2\beta\phi}+
2e^{-\beta\phi}-3\right]~, \nn
\eea
where this time $\beta=\lambda/m^2$.
The special properties of this Lagrangian have been been noticed
by various authors over the years, the
earliest probably being M.Tzitz\'eica, in an article published in
1910.

\exercise{
verify the relationship between the $\phi^3$ and 
$\phi^4$ couplings in the Lagrangian just given by means of a tree-level
calculation.\\[2pt]
%
%\noindent
%{\small\trivlist\item[]
%\item[]
Suggestion: consider a $2\rightarrow 3$ process with both {\it in}\
momenta equal to $(1,1)$, and one of the {\it out}\ momenta equal 
to $(1{+}\delta,(1{+}\delta)^{-1})$ with $\delta$ small. For
the desired result it will suffice to demand that the contributions
to the amplitude
proportional to $\delta^{-2}$ cancel once all relevant diagrams have
been added together. Even this is a little subtle\dots}

One last comment on the uniqueness question: it is easy to see
that all possibilities for a single interacting
massive scalar field with no
tree-level production have now been exhausted. Starting with a
$\phi^3$ or $\phi^4$ interaction term must, if it works at all, 
lead to one of the two theories just discussed: the higher
$\phi^m$ couplings are uniquely determined by the need to
cancel the constant part of the $2\rightarrow m{-}2$ production
amplitude. On the other hand, if both the $\phi^3$ and $\phi^4$
couplings are set to zero, then the same argument shows that all
higher couplings must also be zero, and the theory is free.

%if the first nonzero interaction is
%$\phi^m$ with $m>4$, then the Lagrangian will fail at the first hurdle
%by predicting a constant nonzero amplitude for this same
%$2\rightarrow m{-}2$ process.

\smallskip
To summarise, it appears that in 1+1 dimensions there are some
interacting Lagrangians with the remarkable property that the
resulting field theories have no tree-level particle production.
Araf'eva and Korepin showed, in 1974, that for the sinh-Gordon model
this is also true at one loop. The tree-level result was a sign of
interesting classical behaviour; that it persists to one loop is
evidence that the quantum theory might also be rather special.

Before continuing along this line, I want to return to the
$3\rightarrow 3$ amplitude. Should we
conclude that its connected part is also
zero? Contrary to initial expectations, the answer to this
question is a definite {\it no}. For the $3\rightarrow 3$ process,
it is no longer legitimate to forget about the $i\epsilon$'s.
For the diagrams of type (A), the intermediate particle
can now be on-shell, and when
this happens the $i\epsilon$ must
be retained until all contributing diagrams have been added together.
This is relevant whenever the set of ingoing momenta
is equal to the set of outgoing momenta, and in such situations it turns
out that the final result is indeed nonzero. Thus the connected part
of the $3\rightarrow 3$ amplitude does not vanish, but it does contain an
additional delta-function which enforces the equality
of the  initial and final sets of momenta. We have found a model for 
which, at least at tree level, the connected $3\rightarrow 3$
amplitude violates at certain points
two of the usual assumptions made of an analytic 
S-matrix:

\smallskip
\noindent $\bullet$ it is {\it not} found by crossing the $2\rightarrow 4$
amplitude;
\smallskip

\noindent $\bullet$ it is {\it not} analytic in the residual momenta
once overall momentum conservation has been imposed.%
\footnote{It should be mentioned that developments in the theory of
analytic S-matrices have included a
general understanding of phenomena such as these, 
which are not restricted
to 1+1 dimensions. See, for example, Chandler (1969) and
Iagolnitzer (1973,1978a). The 1+1 dimensional case is treated 
in Iagolnitzer (1978b).}
\smallskip

Even more remarkably, the interaction, while nontrivial, affects 
the participating particles in a minimal way: it does not
change their momenta.
It is clear that something odd is going on, but it is not so
clear quite what, and even less clear why. Evaluating yet more Feynman
diagrams is unlikely to shed much light on these questions, and
besides, an infinite amount of work would be needed before we could be
completely sure that any of these properties feature in the full
quantum theory. A more sophisticated approach is needed. What could force
these amplitudes to vanish, irrespective of the structure of the
Feynman diagrams? One possibility is that conservation laws might
limit the set of {\it out} states accessible from any given {\it in} 
state. The far-reaching consequences of this idea are the subject of 
the next lecture.

\section{Conserved quantities and factorisability}
After the somewhat informal introduction, the time has come to be a
little more precise, at least to the extent of pausing to set up some
notation.

First, I should allow for more than one particle type, so different
masses $m_a$, $m_b$ and so on make an appearance. A single
particle of mass $m_a$ will be on-shell when  its light-cone momenta
$p_a$, $\bar p_a$ satisfy $p_a\bar p_a=m_a^2$. It will be convenient 
to solve this equation not via the variable 
$a=p_a/m_a$ used in the last lecture, but
rather via a parameter $\theta=\log a$ called the rapidity. Thus,
\[
p_a=m_ae^{\theta_a}\quad,\quad
\bar p_a=m_ae^{-\theta_a}~.
\]
Recall
that $a$ was a positive real number for the forward component
of the mass shell; this corresponds to $\theta$
ranging over the entire real axis.  The 
backwards component of the mass shell, 
found by negating $a$, can be parametrised 
by this same rapidity so long as it is shifted onto the line
${\rm Im}\,\theta=\pi$. This will be relevant when discussing
the crossing of amplitudes.

An $n$-particle asymptotic state can now be written as
\[
\ket{A_{a_1}(\theta_1)A_{a_2}(\theta_2)\dots 
A_{a_n}(\theta_n)}_{in\atop out}
\]
where the symbol $A_{a_i}(\theta_i)$ denotes a particle of type $a_i$,
travelling with rapidity $\theta_i$. By smearing the momenta a
little so as to produce wavepackets, each particle can be assigned an 
approximate position at each moment. In a massive theory, the only
sort of theory I will be bothering with, all interactions are short-ranged
and so the state behaves like a collection of free particles except at
times when two or more wavepackets overlap. All of this can be made
more precise, but not in these lectures.

An {\it in} state is characterised by there being no further
interactions as $t\rightarrow-\infty$. This means that the fastest
particle must be on the left, the slowest on the right, with all of
the others ordered in between. It is convenient to represent this
situation by giving the $A_{a_i}(\theta_i)$ a life outside 
the $\ket{~}_{in}$
and $\ket{~}_{out}$ ket vectors, thinking of them as noncommuting
symbols with their order on the page reflecting the spatial ordering
of the particles that they represent. Thus an {\it in} state would be
written
\[
A_{a_1}(\theta_1)A_{a_2}(\theta_2)\dots A_{a_n}(\theta_n)
\]
with
\[
\theta_1>\theta_2>\dots>\theta_n~.
\]

Similarly, an {\it out} state has no further interactions
as $t\rightarrow +\infty$, and so each particle must be to the
left of all particles travelling faster than it, and to the right of
all particles travelling slower. In terms of the non-commuting symbols, 
one such state is
\[
A_{b_1}(\theta_1)A_{b_2}(\theta_2)\dots A_{b_n}(\theta_n)
\]
now with
\[
\theta_1<\theta_2<\dots<\theta_n~.
\]
Products of the symbols with other orderings of the rapidities can be
thought of as representing states at other times when all the particles are 
momentarily well-separated. Asymptotic completeness translates, at 
least partially, into the claim that any such product can be expanded 
either as a sum of products in the {\it in}-state ordering, or as a sum 
of products in the {\it out}-state ordering.

The S-matrix provides the mapping between the {\it in}-state basis
and the {\it out}-state basis. In the new notation this reads, for a
two-particle {\it in}-state,
\[
A_{a_1}(\theta_1)A_{a_2}(\theta_2)=
\sum_{n=2}^{\infty}\,\sum_{\theta'_1{<}\dots {<}\theta'_n\atop{}}\!
S^{b_1\dots b_n}_{~a_1a_2}(\theta_1,\theta_2;\theta'_1\dots
\theta'_n)
A_{b_1}(\theta'_1)\dots A_{b_n}(\theta'_n)~,
\]
where $\theta_1>\theta_2$, a sum on $b_1\dots b_n$ is implied,
and the sum on the $\theta'_i$ will generally involve a number 
of integrals, with the rapidities appearing additionally
constrained by the overall conservation of left- and
right- lightcone momenta:
\[
m_{a_1}e^{\pm\theta_1}+m_{a_2}e^{\pm\theta_2} =
m_{b_1}e^{\pm\theta'_1}+\dots+
m_{b_n}e^{\pm\theta'_n}~.
\]
The notation works because the number of dimensions of space, namely
$1$, matches the `dimensionality' of a sequence of symbols in a
line of mathematics; it can't be used for 
higher-dimensional theories. However, at this stage 
it makes no mention of integrability, and can be set up for
any massive quantum field theory in 1+1 dimensions.%
\footnote{One caveat, though: in nonintegrable theories amplitudes for
the scattering of wavepackets usually depend on impact parameters
as well as momenta. Thus in general the notation should not be taken
too literally, but rather used as a shorthand for recording momentum
space results. In integrable cases we'll see shortly that this
dependence goes away, and so I can afford to be a little careless
about this point.}

Next, to the conserved quantities. One such is
energy-momentum, a spin-one operator.
In lightcone components this acts on a one-particle state as
\[
P\ket{A_a(\theta)}=m_ae^{\theta}\ket{A_a(\theta)}
\quad,\quad
\bar P\ket{A_a(\theta)}=m_ae^{-\theta}\ket{A_a(\theta)}~.
\]
Beyond this, operators can be envisaged transforming in higher
representations of the 1+1 dimensional Lorentz group:
\[
Q_s\ket{A_a(\theta)}=q^{(s)}_ae^{s\theta}\ket{A_a(\theta)}~.
\]
The integer $s$ is called the (Lorentz) spin of $Q_s$. Since $Q_{|s|}$
transforms as $s$ copies of $P$, and $Q_{-|s|}$ as $s$ copies of $\bar
P$, it makes sense to think of $Q_s$ and $Q_{-s}$ as rank $|s|$ 
objects. The simple `left-right' splitting is special to 1+1 dimensions.

I'll only consider those operators $Q_s$ that come as integrals of local
densities, and this has the important consequence that their action on
multiparticle wavepackets is additive:
\[
Q_s\ket{A_{a_1}(\theta)\dots A_{a_n}(\theta_n)}
=(q^{(s)}_{a_1}e^{s\theta_1}+
\dots+q^{(s)}_{a_n}e^{s\theta_n})
\ket{A_{a_1}(\theta)\dots A_{a_n}(\theta_n)}~.
\]

These are called local conserved charges and they
are all in involution (they commute) since, essentially by assumption,
they have been simultaneously diagonalised by the basis of asymptotic
multiparticle states that I have chosen. This is not inevitable:
nonlocal charges, often associated with fractional-spin operators, 
can be very important. The papers of L\"uscher
(1978), Zamolodchikov (1989c) and Bernard and Leclair (1991)
are good starting-points for
those interested in this aspect of the subject.

Even without the more exotic possibilities, the consequences of
the extra local conserved charges are profound. In fact, Coleman
and Mandula showed in 1967 that in three spatial
dimensions the existence of even just one conserved charge
transforming as a tensor of second or higher rank forces the
S-matrix of the model to be trivial. (For a simple-minded explanation
of this fact, see later in this lecture.) This is not true in 1+1
dimensions, but nevertheless the possibilities for the S-matrix
are severely limited: it must be consistent with

\smallskip
\noindent{$\bullet$} no particle production;
\smallskip

\noindent{$\bullet$} equality of the sets of initial and final
momenta;
\smallskip

\noindent{$\bullet$} factorisability of the $n\rightarrow n$ S-matrix
into a product of $2\rightarrow 2$ S-matrices.
\smallskip

The first two of these properties sum up the
behaviour which had emerged experimentally 
by the end of the last lecture, and the third is a
bonus, rendering the task of finding the full S-matrices of a
whole class of 1+1 dimensional models genuinely feasible.

I shall outline a couple of arguments for why these
properties should follow from the existence of the conserved 
charges. 

The first simply imposes the conservation of the
charges directly.
Consider an $n\rightarrow m$ amplitude, with ingoing particles
$A_{a_1}(\theta_1),\dots$, $A_{a_n}(\theta_n)$, and outgoing particles
$A_{b_1}(\theta'_1),\dots$, $A_{b_m}(\theta'_m)$. 
If a charge $Q_s$ is conserved, then
an initial eigenstate of $Q_s$ with a given eigenvalue must
evolve into a superposition of states all sharing that same
eigenvalue. For the amplitude under discussion this implies
that
\[
q^{(s)}_{a_1}e^{s\theta_1}+\dots+q^{(s)}_{a_n}e^{s\theta_n}
=
q^{(s)}_{b_1}e^{s\theta'_1}+\dots+q^{(s)}_{b_n}e^{s\theta'_m}~.
\]
Now if conserved charges $Q_s$ exist for infinitely many values
of $s$, then there will be infinitely many such equations,
and for generic {\it in} momenta the only way to satisfy them all
will be the trivial one, namely $n=m$ and, perhaps after a
reordering of the {\it out} momenta,
\[
\theta_i=\theta'_i\quad;\quad q^{(s)}_{a_i}=q^{(s)}_{b_i}\qquad
i=1\dots n~,
\]
where $s$ runs over the spins of the non-trivial conserved
charges with nonzero spin
(or over all the nonzero integers, if we agree to set $q^{(s)}\equiv
0$ for those $s$ at which a local conserved charge cannot be
defined). This does not quite imply that the outgoing set of labels, 
$\{b_1,\dots b_n\}$, is equal to the ingoing set $\{a_1,\dots a_n\}$
-- they just need to agree about the values of all of the nonzero spin
conserved charges. Nevertheless, it is enough to establish the absence
of particle production, and the equality of the initial and final sets
of momenta, though factorisability is harder to see from this
point of view. One caveat should also be mentioned: in
many models, it turns out that there are some
solutions to the
conservation constraints with $n\neq m$. However these are only found
for exceptional sets of ingoing momenta, which are unphysical to
boot, so this fact does not change the conclusions for the S-matrix.
(In fact, they are associated with solutions to the conserved charged
bootstrap equations, a topic to be discsussed in a later lecture.)
A more severe problem comes with the realisation that this argument
hasn't escaped the infinite workload mentioned at the end of the last
lecture. Consider, for example, a two-particle collision. As the
relative momenta of the incident particles increases, the number of
particles permitted energetically in the out state grows without
limit. To be absolutely sure that, no matter how fast the two
particles are fired at each other, only two particles will come out, 
infinitely
many conservation constraints are needed. This might not matter --
practical considerations are always going to limit the relative
momenta to which we have access -- were we not ambitious enough
to hope for an {\it exact} formula for the S-matrix. This requires an
understanding of all energy scales, and so the infinite amount of work
appears to be unavoidable. 

This should be motivation enough for the second argument, which can be
found in a 1980 article by Parke, itself building on an observation
which dates back at least to Shankar and Witten (1978). 
The argument also establishes factorisability and imposes the
Yang-Baxter equation on the two-particle S-matrix. The key is to make
use of the fact that we're dealing with a local, causal quantum field 
theory, by considering the effect of the conserved charges on
localised wavepackets.

First take a single-particle state, with position space wavefunction
\[
\psi(x)\propto \int^{\infty}_{-\infty}\!dp
e^{-a^2(p-p_1)^2}e^{ip(x-x_1)}.
\]
This describes a particle with spatial momentum approximately $p_1$,
and position approximately $x_1$. Act on this with an operator
giving a momentum-dependent phase factor $e^{-i\phi(p)}$. The
wavefunction becomes
\[
\tilde\psi(x)\propto \int^{\infty}_{-\infty}\!dp
e^{-a^2(p-p_1)^2}e^{ip(x-x_1)}e^{-i\phi(p)}.
\]
Most of the integral comes from $p\approx p_1$, and $\phi(p)$ can be
expanded in powers of $(p{-}p_1)$ to find $\tilde p_1$ and $\tilde
x_1$, the revised values of the momentum and position:
\[
\tilde p_1=p_1\qquad,\qquad \tilde x_1=x_1+\phi'(p_1)~.
\]

For a multiparticle state a product of one-particle wavefunctions will
be a good approximation when the particles are well separated, and on
such a state $\ket{p_ap_b\dots}$, the action is to shift the position
of particle $a$ by $\phi'(p_a)$, that of $b$ by $\phi'(p_b)$, and so
on.

Strictly speaking, for compatibility with the earlier discussions I
should now consider the actions of the operators $Q_{|s|}$ and
$Q_{-|s|}$, as Parke did in his article. However the essentials of the
argument will be conveyed if I instead assume 
the conservation of operators $P_s$ acting on one-particle and
well-separated multiparticle states as $(P_1)^s$, with $P_1$ the
spatial part of the two-momentum operator. Acting with
$e^{-i\alpha P_s}$, the phase factor is $\phi_s(p)=\alpha p^s$, so a
particle with momentum $p_a$ will have its position shifted by
$s\alpha p_a^{s-1}$. The case $s=1$, momentum itself, just translates
every particle by the same amount $\alpha$. But, crucially, for $s>1$
particles with different momenta are moved by different amounts.

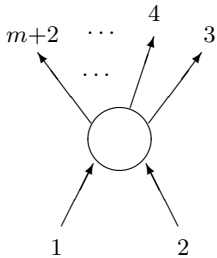
\begin{figure}
\begin{picture}(300,120)(-65,-45)
\thinlines
\put(92,-33){\vector(1,2){12}}
\put(136,-33){\vector(-1,2){12}}
\put(125,6){\vector(3,4){20}}
\put(118,12){\vector(1,3){9}}
\put(103,6){\vector(-3,4){20}}
\put(114,0){\circle{24}}
\put(90,-38){\makebox(0,0)[t]{$1$}}
\put(138,-38){\makebox(0,0)[t]{$2$}}
\put(81,37){\makebox(0,0)[b]{$m{+}2$}}
\put(148,37){\makebox(0,0)[b]{$3$}}
\put(127,45){\makebox(0,0)[b]{$4$}}
\put(105,24){\makebox(0,0)[b]{$\dots$}}
\put(107,40){\makebox(0,0)[b]{$\dots$}}
\end{picture}
\caption[ ]{A $2\rightarrow m$ process}
\label{ttm}
\end{figure}

The argument continues as follows. First consider a $2\rightarrow m$
process, labelled as in figure~\ref{ttm}.

For the amplitude to be non-vanishing, the
time when the first two particles collide, call it $t_{12}$, 
must precede the time $t_{23}$ when the trajectory of particle $2$,
the slower incomer, intersects that of particle $3$, the fastest
outgoer:
\[
t_{12}\leq t_{23}~.
\]
Why should this be so? Nothing can happen until the wavepackets of
particles $1$ and $2$ overlap. After this, it suffices to follow the
path of the rightmost particle until all have separated in order to
establish the inequality.  Note that this could be violated on
microscopic timescales, but not macroscopically: hence the 
term `macrocausality' for this sort of property.

The constraint is rendered vastly more powerful if there is a
conserved higher-spin charge $P_s$ in the model. Since it must commute
with the S-matrix, we have
\[
\bra{final}S\ket{initial}=
\bra{final}e^{i\alpha P_s}Se^{-i\alpha P_s}\ket{initial}
\]
and so $e^{i\alpha P_s}$ can be used to rearrange the initial and 
final configurations without
changing the amplitude. All that remains -- and for this you should
consult Parke's article -- is to show that if any of the outgoing
rapidities are different from $\theta_1$ or $\theta_2$, then shifting the
configurations around in this way will give a pattern of trajectories
for which $t_{12}>t_{23}$. By macrocausality the amplitude for this
pattern must vanish, and then by the insensitivity of the amplitudes to
shifts induced by $e^{i\alpha P_s}$ all of the other
amplitudes, including the one initially under consideration, must also
vanish. Hence the only possibilities for the two incoming particles
are two outgoing particles with the same pair of rapidities as before
the interaction, which is the result required for $n{=}2$.

To complete the missing step, Parke actually needed to assume the
existence of {\it two} extra charges of higher spin. However, since a
parity-conjugate pair $Q_s$, $Q_{-s}$ will do, this is scarcely a
problem, at least in parity-symmetric theories.

For completeness, I should mention that there is a quicker argument
for this $2\rightarrow m$ amplitude, to be
found in Polyakov (1977), which revives the previous line of
reasoning, though with a slight twist. 
As previously noted, if the first argument is attempted
with the time in figure~\ref{ttm} running up
the page, more and more conserved charges will be needed as $m$
increases in order to eliminate all the undesired possibilities for
the final configuration. But by $T$-invariance, the $2\rightarrow m$
amplitude will only be nonvanishing if the same is true of the
time-reversed $m\rightarrow 2$ amplitude. But now there are just two
outgoing momenta, and these are fixed, 
up to a discrete ambiguity, by energy-momentum
conservation and the on-shell condition.
After this, {\it any} extra charge will suffice to
eliminate the process. Economical as this argument is, it does not
cover the general $m\rightarrow n$ amplitude,
and factorisability and the Yang-Baxter equation are
missed.

One other aside before moving on: 
however the higher-spin conserved
charges are used to reshuffle the
positions of an incident pair of particles, if their rapidities differ
then their trajectories will still cross somewhere. This is special
to 1+1 dimensions: with more than
one spatial dimension to play with, conserved higher
spin charges can be used to make
trajectories miss each other completely, even on macroscopic scales. 
It is then but a short step to deduce that
the S-matrix must be trivial -- and this, in
admittedly sketchy form, is an argument for the Coleman-Mandula theorem
alluded to earlier.

\begin{figure}
\begin{picture}(300,110)(0,-25)
\thinlines
\put(30,0){\begin{picture}(100,100)(0,0)
\put(0,0){\vector(3,4){57}}
\put(15,-5){\vector(0,1){90}}
\put(57,4){\vector(-3,4){57}}
\put(15,20){\circle*{4}}
\put(15,60){\circle*{4}}
\put(30,40){\circle*{4}}
\put(27,-25){$(1)$}
\end{picture}}
\put(140,0){\begin{picture}(100,100)(0,0)
\put(3,4){\vector(3,4){54}}
\put(30,-3){\vector(0,1){86}}
\put(57,4){\vector(-3,4){54}}
\put(30,40){\circle*{5}}
\put(25,-25){$(2)$}
\end{picture}}
\put(250,0){\begin{picture}(100,100)(0,0)
\put(3,4){\vector(3,4){57}}
\put(45,-5){\vector(0,1){90}}
\put(60,0){\vector(-3,4){57}}
\put(45,20){\circle*{4}}
\put(45,60){\circle*{4}}
\put(30,40){\circle*{4}}
\put(25,-25){$(3)$}
\end{picture}}
\end{picture}
\caption[ ]{Possibilities for a $3\rightarrow 3$ process}
\label{ttt}
\end{figure}
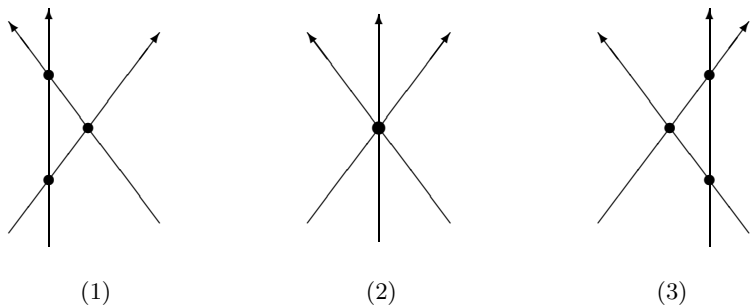

To deal with three incoming particles, consider first how the
trajectories would look were there no interactions in the model.
Figure~\ref{ttt} shows the three distinct possibilities -- which one
actually occurs depends on the particular spatial positions of the
incident wavepackets.

In cases 1 and 3, when we switch the interaction back on again the
results just established for two incident particles, together with
locality, are enough to see that the pictures do not change in any
essential way. Furthermore, as the interaction proceeds by a series of
two-body collisions, these amplitudes must {\it factorise} into
products of $2\rightarrow 2$ amplitudes. Case 2 in general would
give
something new. However, using $e^{-i\alpha P_s}$ in the manner
discussed at length above, it can be converted into one of the other
cases. Hence there is never any particle production,
individual momenta are conserved, and the amplitudes always factorise.
In addition, the equality of amplitudes 1 and 3 gives a constraint on
the two-body amplitudes, known as the Yang-Baxter equation.
More on this in the next lecture, once the necessary notation has been
set up.

To go beyond three incoming particles, an inductive argument can be used, 
showing that a set of $n$
incident particles can always be shuffled around in such a way that
the interaction occurs via a sequence of events in which at most
$n{-}1$ particles are participating. 

The ultimate conclusion is that in any local scattering theory in 1+1 
dimensions with a couple of local higher-spin conserved charges (and a
parity-conjugate pair $\{Q_s,Q_{-s}\}$, $s>1$, will certainly do),
there is no particle production, the final set of momenta is equal to
the initial set, and the $n\rightarrow n$ S-matrix factorises into a
product of $2\rightarrow 2$ S-matrices. These are the three
properties promised earlier, and now they
can be established with only a finite amount of work.

Finally, I would like to mention a mild paradox that might at first sight
seem troubling.  If $\{p'_1\dots p'_n\}=\{p_1\dots p_n\}$ for every
set of ingoing momenta, then surely $\sum (p_a)^s$ is conserved for
all $s$, resulting in conserved charges at {\it all} spins, in any
model for which the arguments above apply? This reasoning misses a key
feature of the objects we are dealing with:
for $Q_s$ to qualify as a local conserved charge, it must be possible
to write it as the integral of a local conserved density:
\[
Q_s=\int^{\infty}_{-\infty}T_{s+1}dx~.
\]
There is no a priori
reason why such a density should exist, even if the sums
$\sum (p_a)^s$ happen to be conserved.
In fact, the set of spins $s$ at which this can
be done forms a rather good fingerprint for a model, and turns out to
constrain its behaviour in important ways. 

\section{The two-particle S-matrix}
Once the two-particle S-matrix is known, factorisability tells us that
the entire S-matrix follows. To find the two-particle
S-matrix becomes the main goal. In the algebraic notation of the last
lecture, we can write
\[
\ket{A_i(\theta_1)A_j(\theta_2)}_{in}=S^{kl}_{ij}(\theta_1-\theta_2)
\ket{A_k(\theta_1)A_l(\theta_2)}_{out}
\]
as
\[
A_i(\theta_1)A_j(\theta_2)=S^{kl}_{ij}(\theta_1-\theta_2)
A_l(\theta_2)A_k(\theta_1)~,
\]
with $\theta_1>\theta_2$ to ensure that {\it in} and {\it out} states
are correctly represented. A sum over $k$ and $l$ is implied, with
$k\neq i$ and $l\neq j$ being possible in those situations where some
particles are not distinguished by the $Q_{s\neq 0}$ conserved
charges. Lorentz boosts
shift rapidities by a constant, and so $S$ only depends on the
difference $\theta_1-\theta_2=\theta_{12}$. 

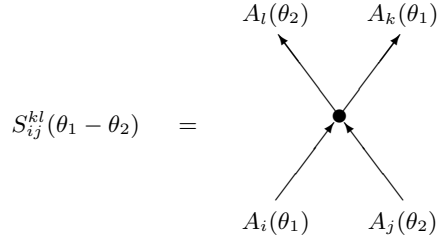
\begin{figure}
\begin{picture}(300,80)(-100,-45)
\thinlines
\put(-10,-6){$S^{kl}_{ij}(\theta_1-\theta_2)~~~~=$}
\put(90,-32){\vector(3,4){22}}
\put(138,-32){\vector(-3,4){22}}
\put(114,0){\vector(3,4){23}}
\put(114,0){\vector(-3,4){23}}
\put(114,0){\circle*{5}}
\put(90,-36){\makebox(0,0)[t]{$A_i(\theta_1)$}}
\put(138,-36){\makebox(0,0)[t]{$A_j(\theta_2)$}}
\put(90,36){\makebox(0,0)[b]{$A_l(\theta_2)$}}
\put(138,36){\makebox(0,0)[b]{$A_k(\theta_1)$}}
\end{picture}
\caption[ ]{The two-particle S-matrix}
\label{tps}
\end{figure}

\vspace{-5pt}
In a theory with $r$ different particle types, knowledge of the $r^4$
functions $S^{kl}_{ij}(\theta)$ will thus give the full S-matrix.
Not all of these functions are independent, and their analytic
properties are heavily constrained. Such general features are the 
subject of this lecture.

First, as just mentioned, in an integrable model the matrix
element $S^{kl}_{ij}$ can only be nonzero if 
$A_i$ and $A_k$, and $A_j$ and $A_l$, agree
on the values of all of the local conserved
charges with nonzero spin
(which, in particular, requires $m_i=m_k$ and $m_j= m_l$).
Next, 
the assumptions of $P$, $C$ and $T$ invariance imply
\[
S^{kl}_{ij}(\theta)=S^{lk}_{ji}(\theta)\quad;\quad
S^{kl}_{ij}(\theta)=S^{\bar k\bar l}_{\bar\imath\bar\jmath}(\theta)\quad;\quad
S^{kl}_{ij}(\theta)=S^{ji}_{lk}(\theta)~.
\]
Analytic properties of the S-matrix are usually discussed in terms of
the Mandelstam variables $s$, $t$ and $u$:
\[
s=(p_1{+}p_2)^2~,\quad
t=(p_1{-}p_3)^2~,\quad
u=(p_1{-}p_4)^2~,
\]
with $s{+}t{+}u=\sum_{i=1}^4m_i^2$. In 1+1 dimensions only one of
these is independent, and it is standard to focus on $s$, the square of
the forward-channel momentum. In terms of the rapidity difference
$\theta_{12}=\theta_1-\theta_2\,$,
\[
s=m_i^2+m_j^2+2m_im_j\cosh\theta_{12}~.
\]
For a physical process, $\theta_{12}$ is real and so $s$ is
real and satisfies $s\ge (m_i{+}m_j)^2$. But we can consider the
continuation of $S(s)$ up into the complex plane. Placing the branch
cuts in the traditional way, this results in a function with the
following properties:

\noindent{$\bullet$} $S$ is a singlevalued, meromorphic function on
the complex plane with cuts on the portions of the real axis $s\le
(m_i-m_j)^2$
and $s\ge (m_i{+}m_j)^2$. Physical values of $S(s)$ are found for $s$
just above the right-hand cut. This first sheet of the full Riemann
surface for $S$ is called the physical sheet.

\noindent{$\bullet$} $S$ is real-analytic: it takes
complex-conjugate values at complex-conjugate points:
\[
S^{kl}_{ij}(s^*)=\left[S^{kl}_{ij}(s)\right]^*.
\]
In particular $S(s)$ is real if $s$ is real and $(m_i{-}m_j)^2\le s\le
(m_i{+}m_j)^2$.

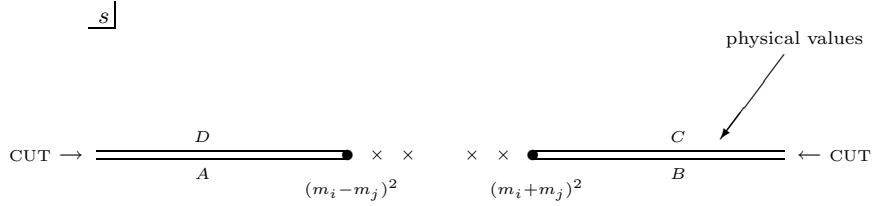
\begin{figure}[ht]
\begin{picture}(300,120)(-45,-50)
\thinlines
\put(-5,0){\makebox(0,0)[r]{$\scriptstyle {\rm CUT}$~$\rightarrow$}}
\put(265,0){\makebox(0,0)[l]{$\leftarrow$~$\scriptstyle {\rm CUT}$}}
\put(0,1.5){\line(1,0){95}}
\put(0,-1.5){\line(1,0){95}}
\put(95,0){\circle*{4}}
\put(165,0){\circle*{4}}
\put(165,1.5){\line(1,0){95}}
\put(165,-1.5){\line(1,0){95}}
\put(95,-9){\makebox(0,0)[t]{$\scriptstyle~(m_i{-}m_j)^2$}}
\put(165,-9){\makebox(0,0)[t]{$\scriptstyle~(m_i{+}m_j)^2$}}
\put(40,5){\makebox(0,0)[b]{$\scriptstyle D$}}
\put(40,-5){\makebox(0,0)[t]{$\scriptstyle A$}}
\put(220,5){\makebox(0,0)[b]{$\scriptstyle C$}}
\put(220,-5){\makebox(0,0)[t]{$\scriptstyle B$}}
\put(260,38){\vector(-3,-4){24}}
\put(264,42){\makebox(0,0)[b]{\scriptsize physical values}}
\multiput(112,0)(36,0){2}
{\makebox(0,0){$\scriptstyle\times~~\times$}}
\put(5,50){\makebox(0,0)[br]{$s$}}
\put(7,48){\line(-1,0){10}}
\put(7,48){\line(0,1){10}}
\end{picture}
\caption[ ]{The physical sheet}
\label{physs}
\end{figure}

The situation is depicted in figure~\ref{physs}.

Unitarity requires that $S(\sP)S^{\dagger}(\sP)=1$
whenever $\sP$ is a physical value for $s$, just
above the right-hand cut: $\sP=s+i0$, $s>(m_i+m_j)^2$.
This should be understood as a matrix
equation, with a sum over a complete set of asymptotic states hiding
between $S$ and $S^{\dagger}$. As $\sP$ grows, it becomes
energetically possible for states with more and more particles to
participate in the sum. Generally this brings the $2\rightarrow m$ 
S-matrix elements into the story with $m=3,4,\dots\,$, and gives
the $2\rightarrow 2$ S-matrix elements a series of branch
points along the real axis, located at the $3,4,\dots$ 
particle thresholds. However for an integrable model these production
amplitudes should all be zero, and so for all physical $\sP$
unitarity reads
\[
S^{kl}_{ij}(\sP)\left[S^{nm}_{kl}(\sP)\right]^*=
\delta^n_i\delta^m_l\,.
\]
With the help of real analyticity this can be rewritten as
\[
S^{kl}_{ij}(\sP)S^{nm}_{kl}(\sM)=\delta^n_i\delta^m_l\,,
\]
with $\sM=s-i0$, just below the right-hand cut. This equation shows
the need for a branch cut running rightwards from the two-particle
threshold $s=(m_i+m_j)^2\,$;
if we accept that the cut actually starts at this threshold, then it is
easy to see that the branch is of square-root type. The argument goes
as follows. Let
$S_{\gamma}(s)$ be the function obtained by analytic continuation of
$S(s)$ once anticlockwise around the branch point. Unitarity amounts
to the requirement that $S(\sP)S_{\gamma}(\sP)=1$ for all physical
values of $\sP$. When written in this way, the 
relation can be analytically continued to all $s$, so
\[
S_{\gamma}(s)=S^{-1}(s)~.
\]
In particular, if $\sM$ is a point just below the cut, then
\[
S_{\gamma}(\sM)=S^{-1}(\sM)=S(\sP)~,
\]
the last equality following from a second application of unitarity.
Now $S_{\gamma}(\sM)$ is just the analytic continuation of $S(\sP)$
twice around $(m_i+m_j)^2$. Therefore, twice round the branch
point gets us back to where we started, and the singularity
is indeed a square root.

So much for the right-hand cut. The left-hand half of the figure, 
containing the second cut running in the opposite direction, can be
understood via
the fundamentally relativistic property of crossing. If one of
the incoming particles, say $j$, is crossed to become outgoing while
simultaneously one of the outgoers, say $l$, crosses in the opposite
sense and becomes ingoing, then the amplitude for another physical
two-particle scattering process results. For this new amplitude the
incomers are $i$ and $\bar l$, and the outgoers $k$ and $\bar\jmath$,
where an overbar has been introduced to denote the (possibly trivial)
operation of conjugation on particle labels. All of this amounts to
looking at figure~\ref{tps} from the side, with the forward-channel
momentum now not $s$ but rather $t=(p_1-p_3)^2$. In this
particular case $p_3=p_2$, and the relation between $t$ and $s$ is 
very simple:
\[
t=(p_1-p_2)^2=2p_1^2+2p_2^2-(p_1+p_2)^2=2m_i^2+2m_j^2-s~.
\]
Crossing symmetry states that the amplitude for this process can be
obtained by analytic continuation of the previous amplitude into a
region of the $s$ plane where $t$ becomes physical, that is $t\in\bbbr$
and $t\ge(m_i+m_j)^2$. Physical amplitudes correspond to approaching
this line segment from above in the $t$ plane, and hence from below in
the $s$ plane. Thus the amplitudes are on the lower edge of the
left-hand cut, marked $A$ on figure~\ref{physs}. In equations:
\bea
S^{kl}_{ij}(\sP)&=&S^{k\bar\jmath}_{i\bar l}(2m_i^2+2m_j^2-\sP)~.\nn\\
\makebox[0pt]{$\uparrow$}~~~~&&
\qquad\qquad\qquad\makebox[0pt]{$\uparrow$}\nn\\
\makebox[0pt]{(on $C$)}~~~~&&
\qquad\qquad\qquad\makebox[0pt]{(on $A$)} \nn
\eea
Clearly the cross-channel branch point at $(m_i-m_j)^2$ must also be
a square root, but this does {\it not} mean that the Riemann surface
for $S(s)$ has just two sheets. Continuing through the left-hand cut
can, and usually does, connect with a different sheet from that found
through the right-hand cut. Stepping up and down to left and
right, the typical $S(s)$, even for an integrable model, lives on an
infinite cover of the physical sheet. 

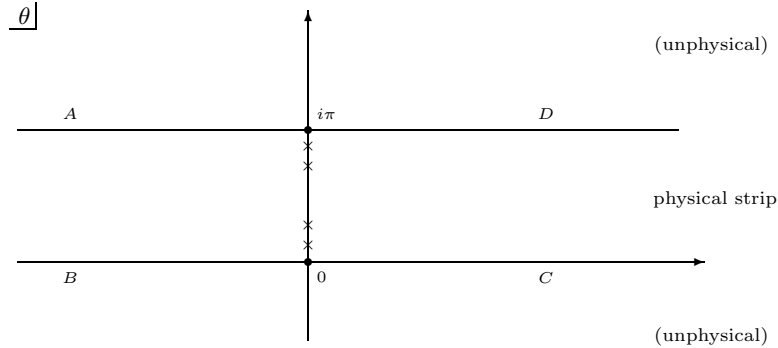
\begin{figure}
\begin{picture}(300,145)(-40,-40)
\thinlines
\put(0,0){\vector(1,0){260}}
\put(110,0){\circle*{3}}
\put(0,50){\line(1,0){250}}
\put(110,50){\circle*{3}}
\put(110,-30){\vector(0,1){125}}
\put(112,-4){\makebox(0,0)[tl]{$\scriptstyle\,0$}}
\put(112,54){\makebox(0,0)[bl]{$\scriptstyle\,i\pi$}}
\put(20,54){\makebox(0,0)[b]{$\scriptstyle A$}}
\put(20,-4){\makebox(0,0)[t]{$\scriptstyle B$}}
\put(200,54){\makebox(0,0)[b]{$\scriptstyle D$}}
\put(200,-4){\makebox(0,0)[t]{$\scriptstyle C$}}
\put(264,22){\makebox(0,0)[b]{\scriptsize physical strip}}
\put(264,80){\makebox(0,0)[b]{\scriptsize (unphysical)~}}
\put(264,-30){\makebox(0,0)[b]{\scriptsize (unphysical)~}}
\multiput(110,10)(0,30){2}
{\makebox(0,0){\shortstack{$\scriptstyle\times$\\$\scriptstyle\times$}}}
\put(5,90){\makebox(0,0)[br]{$\theta$}}
\put(7,88){\line(-1,0){10}}
\put(7,88){\line(0,1){10}}
\end{picture}
\caption[ ]{The $\theta$ plane}
\label{physst}
\end{figure}

This looks rather complicated, but simplifies considerably if,
following Zamolodchikov, attention is switched
from the Mandelstam variable $s$ to
the rapidity difference $\theta$. The transformation is
\bea
\theta&=&\cosh^{-1}\left(\frac{s-m_i^2-m_j^2}{2m_im_j}\right)\nn\\
      &=&\log\left[\frac{1}{2m_im_j}\left(s-m_i^2-m_j^2 +
\sqrt{(s-(m_i{+}m_j)^2)(s-(m_i{-}m_j)^2)}\right)\right] \nn
\eea
and it maps the physical sheet into the region
\[
0\le\im\theta\le\pi
\]
of the $\theta$ plane called the physical strip. Most importantly, the
cuts are opened up, so that $S(\theta)$ is analytic at the images $0$
and $i\pi$ of the two physical-sheet branch points, and also at the
images $i n\pi$ of the branch points on all of the other, unphysical,
sheets. Since, by integrability, these are expected to be the only
branch points, $S$ is a meromorphic function of $\theta$. The other
sheets are mapped onto a succession of strips 
\[
n\pi\le\im\theta\le (n{+}1)\theta~.
\]
The new image of the Riemann surface is shown in figure
\ref{physst}.

The previous relations can now be translated to give a list of
constraints on $S(\theta)$ to be carried forward into later
lectures:

\smallskip
\noindent $\bullet$ Real analyticity: $S(\theta)$ is real for $\theta$
purely imaginary;
\smallskip

\noindent $\bullet$ Unitarity:\quad $S^{nm}_{ij}(\theta)S^{kl}_{nm}(-\theta)=
\delta^k_i\delta^l_j\,;$
\smallskip

\noindent $\bullet$ Crossing:\quad\, $S^{kl}_{ij}(\theta)=
S^{k\bar\jmath}_{i\bar l}(i\pi-\theta)\,.$
\smallskip

\noindent
A couple of remarks: first, both the unitarity and the crossing equations
can now be analytically continued, and apply to the whole of the
$\theta$ plane, not just along the line segments of physical values.
Second, the unitarity constraint means that it is consistent to extend
the algebraic relation
\[
A_i(\theta_1)A_j(\theta_2)=S^{kl}_{ij}(\theta_1-\theta_2)
A_l(\theta_2)A_k(\theta_1)
\]
to $\theta_1<\theta_2$.
Unitarity then becomes a consequence of the algebra, and the
single-valued nature of products of the non-commuting symbols.

Finally to some unfinished business from the previous lecture.
Shifting trajectories showed that the amplitudes (1) and (3) of figure
\ref{ttt} must be equal. If the two-particle S-matrix is not
completely diagonal, this equality is not automatic but instead
results in the following consistency condition:
\[
S^{\beta\alpha}_{ij}(\theta_{12})
S^{n\gamma}_{\beta k}(\theta_{13})
S^{ml}_{\alpha\gamma}(\theta_{23})=
S^{\beta\gamma}_{jk}(\theta_{23})
S^{\alpha l}_{i\gamma}(\theta_{13})
S^{nm}_{\alpha\beta}(\theta_{12})~,
\]
where $\theta_{ab}=\theta_a-\theta_b$, and $\theta_1$, $\theta_2$ and
$\theta_3$ are the rapidities of particles $i$, $j$ and $k$. This is
the Yang-Baxter equation, forced by the ability of the conserved
charges to shift particle trajectories around. In theories where
particles appear in multiplets transforming under some symmetry group,
this equation together with some minimality assumptions is often
enough to conjecture the complete functional form of $S$. 
The equation is equivalent
to associativity for the algebra of the $A_i(\theta)$'s: moving from
$A_i(\theta_1)A_j(\theta_2)A_k(\theta_3)$ to a sum of products
$A_l(\theta_3)A_m(\theta_2)A_n(\theta_1)$, the result is independent
of the order of the pair transpositions, if and only if the
Yang-Baxter equation holds for the two-particle S-matrix elements.
\goodbreak

\section{Pole structure and bound states}
The remaining features of figures \ref{physs} and
\ref{physst} are the crosses marked between the two thresholds. 
The first things one might expect to find in these locations
are simple poles
corresponding to stable bound states, appearing either in the
forward ($s$) or the crossed ($t$) channel:

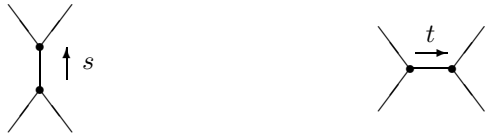
\begin{picture}(300, 70)(0,10)
\thinlines
\put( 82,36){\circle*{3}}
\put( 82,52){\circle*{3}}
\put(70,20){\line( 3, 4){ 12}}
\put(82,36){\line( 3,-4){ 12}}
\put(82,36){\line( 0, 1){ 16}}
\put(92,40){\vector( 0, 1){ 12}}
\put(98,44){$s$}
\put(70,68){\line( 3,-4){ 12}}
\put(82,52){\line( 3, 4){ 12}}
\put(238,44){\circle*{3}}
\put(222,44){\circle*{3}}
\put(210,28){\line( 3, 4){ 12}}
\put(210,60){\line( 3,-4){ 12}}
\put(222,44){\line( 1, 0){ 16}}
\put(224,50){\vector( 1, 0){ 12}}
\put(228,54){$t$}
\put(238,44){\line( 3,-4){ 12}}
\put(238,44){\line( 3, 4){ 12}}
\end{picture}

This is potentially important -- for example, it might signal the
presence of hitherto unsuspected particles in the spectrum of the
model. Most of the remaining lectures will be spent on
this point. I'll start by recalling a selection of reasons why 
the association between simple poles in an S-matrix and bound states
is natural:

\smallskip\noindent
$\bullet$ Potential scattering: in quantum mechanics, if the S-matrix
for the scattering of a particle off a potential has a pole -- which,
as it happens, is always simple -- then it is possible to use it to
{\it construct} a wavefunction for the particle bound to the
potential;

\smallskip\noindent
$\bullet$ tree-level Feynman diagrams;

\smallskip\noindent
$\bullet$ an `axiom', justified if by nothing else by experience in 3+1
dimensions. 
\smallskip

It turns out that life isn't so simple in 1+1 dimensions. To
explain this I'll use the grandparent of all integrable
field theories, the sine-Gordon model. Take the sinh-Gordon Lagrangian
introduced in the first lecture, and replace $\beta$ by $i\beta$.
Re-zeroing the energy of the classical ground state, you will find
\[
{\cal L}=\frac{1}{2}(\partial\phi)^2-V(\phi)
\]
with
\[
V(\phi)=\frac{m^2}{\beta^2}\left[1-\cos(\beta\phi)\right]~.
\]
There is extra structure here, as compared to the sinh-Gordon model,
since there are infinitely-many classical vacua, 
$\phi(x)=2n\pi/\beta$,
$n\in\bbbz$. There is a conserved, spin-zero topological charge $Q_0$:
\[
Q_0=\frac{\beta}{2\pi}\int^{\infty}_{-\infty}\partial_x\phi dx
\]
which is non-zero for configurations which
interpolate between different vacua. (Formally the charge can also be
defined, and is conserved, for the sinh-Gordon model -- the only
problem is that it is identically zero for all finite-energy
configurations.) 

Classically, the model has a soliton $s$ with $Q_0=+1$, and an
antisoliton $\bar s$ with $Q_0=-1$, both with mass $M$, say, and both
interpolating between neighbouring vacua. There are no classically
stable solutions with $|Q_0|>1$. Solitons repel solitons, antisolitons
repel antisolitons, but solitons and antisolitons attract. Therefore
the classical theory additionally sees a continuous family of so-called
breather solutions, which are $s\bar s$ bound states. 
Although not static, they are periodic in time and
in most respects behave just like further particle states. Their
`masses' range from $0$ (tightly-bound) to $2M$ (almost unbound).

In the quantum theory, the breather spectrum becomes discrete, just as
would be expected from quantum mechanics. If $s$ and $\bar s$ have mass
$M$, then the breather masses are
\[
M_k=2M\sin\frac{\pi k}{h}~,\qquad k=1,2,\dots<\frac{8\pi}{\beta^2}-1
\]
where
\[
h=\frac{16\pi}{\beta^2}\left(1-\frac{\beta^2}{8\pi}\right)~.
\]
This was found by Dashen, Hasslacher and Neveu in 1975 via a semiclassical
quantisation of the two-soliton solution, and is thought to be exact. Notice
that as $\beta\rightarrow 0$, corresponding to the classical limit,
the continuous breather spectrum is recovered.

The S-matrix elements of the solitons provide an illustration of the
notational technology set up earlier. The model turns out to possess
higher-spin conserved charges, and so all of the previous discussions
apply. However at generic values of $\beta^2$ none of them
breaks the $\phi\rightarrow -\phi$ symmetry of the original Lagrangian, 
and so none can be used to distinguish the soliton from the
antisoliton. That leaves $Q_0$, which makes a fine job of
distinguishing a single soliton from a single antisoliton
but, as we shall now see, is not quite powerful
enough when acting on two-particle states to rule out nondiagonal
scattering. 

Consider a general two-particle in-state
$\ket{A(\theta_1)_{s,\bar s}A(\theta_2)_{s,\bar s}}_{in}\,$, 
each particle either a
soliton or an antisoliton. The higher-spin charges can be used in the
ways explained earlier to show
that any out-state into which this state evolves must again contain two
particles with rapidities $\theta_1$ and $\theta_2$, 
each either a soliton or an antisoliton. Thus before recourse is made
to the spin-zero charge, a four-dimensional space of out-states is
available. The topological charge $Q_0$ acts in this space as
follows:
\[
Q_0
\left(\begin{array}{c}
\ket{A_s(\theta_1)A_s(\theta_2)}\\
\ket{A_s(\theta_1)A_{\bar s}(\theta_2)}\\
\ket{A_{\bar s}(\theta_1)A_s(\theta_2)}\\
\ket{A_{\bar s}(\theta_1)A_{\bar s}(\theta_2)}
\end{array}\right)  =
\left(\begin{array}{cccc}
{}~2~&&&\\
&~0~&&\\
&&~0~&\\
&&&-2~
\end{array}\right)
\left(\begin{array}{c}
\ket{A_s(\theta_1)A_s(\theta_2)}\\
\ket{A_s(\theta_1)A_{\bar s}(\theta_2)}\\
\ket{A_{\bar s}(\theta_1)A_s(\theta_2)}\\
\ket{A_{\bar s}(\theta_1)A_{\bar s}(\theta_2)}
\end{array}\right)~.
\]
The soliton-soliton and antisoliton-antisoliton states are
picked out uniquely, and therefore must scatter diagonally. 
The same cannot be said for the remaining two states, and it is
through this loophole that
nondiagonal scattering enters the story. 

Taking charge conjugation symmetry into account, there are just three
independent amplitudes to be determined. With
$\theta=\theta_1-\theta_2$, these can be written as:
\[
\left(\begin{array}{c}
\ket{A_s(\theta_1)A_s(\theta_2)}_{in}\\
\ket{A_s(\theta_1)A_{\bar s}(\theta_2)}_{in}\\
\ket{A_{\bar s}(\theta_1)A_s(\theta_2)}_{in}\\
\ket{A_{\bar s}(\theta_1)A_{\bar s}(\theta_2)}_{in}
\end{array}\right)  =
\left(\begin{array}{cccc}
{}S(\theta)&&&\\
&S_T(\theta)&S_R(\theta)&\\
&S_R(\theta)&S_T(\theta)&\\
&&&S(\theta)
\end{array}\right)
\left(\begin{array}{c}
\ket{A_s(\theta_1)A_s(\theta_2)}_{out}\\
\ket{A_s(\theta_1)A_{\bar s}(\theta_2)}_{out}\\
\ket{A_{\bar s}(\theta_1)A_s(\theta_2)}_{out}\\
\ket{A_{\bar s}(\theta_1)A_{\bar s}(\theta_2)}_{out}
\end{array}\right)~.
\]
The same information can be given pictorially:

\begin{picture}(300,40)(-50,-7)
\thinlines
\put(30,8){$\displaystyle S(\theta)~~=$}
\put(85,0){\line(1,1){20}}
\put(105,0){\line(-1,1){20}}
\put(95,10){\circle*{5}}
\put(84,23){$\scriptstyle s$}
\put(104,23){$\scriptstyle s$}
\put(83.5,-6){$\scriptstyle s$}
\put(103.5,-6){$\scriptstyle s$}
\put(125,8){$=$}
\put(155,0){\line(1,1){20}}
\put(175,0){\line(-1,1){20}}
\put(165,10){\circle*{5}}
\put(154,23){$\scriptstyle\bar s$}
\put(174,23){$\scriptstyle\bar s$}
\put(153.5,-6.5){$\scriptstyle\bar s$}
\put(173.5,-6.5){$\scriptstyle\bar s$}
\end{picture}

\begin{picture}(300,50)(-50,-7)
\thinlines
\put(25,8){$\displaystyle S_T(\theta)~~=$}
\put(85,0){\line(1,1){20}}
\put(105,0){\line(-1,1){20}}
\put(95,10){\circle*{5}}
\put(84,23){$\scriptstyle s$}
\put(104,23){$\scriptstyle\bar s$}
\put(83.5,-6.5){$\scriptstyle\bar s$}
\put(103.5,-6){$\scriptstyle s$}
\put(125,8){$=$}
\put(155,0){\line(1,1){20}}
\put(175,0){\line(-1,1){20}}
\put(165,10){\circle*{5}}
\put(154,23){$\scriptstyle\bar s$}
\put(174,23){$\scriptstyle s$}
\put(153.5,-6){$\scriptstyle s$}
\put(173.5,-6.5){$\scriptstyle\bar s$}
\end{picture}

\begin{picture}(300,50)(-50,-7)
\thinlines
\put(25,8){$\displaystyle S_R(\theta)~~=$}
\put(85,0){\line(1,1){20}}
\put(105,0){\line(-1,1){20}}
\put(95,10){\circle*{5}}
\put(84,23){$\scriptstyle\bar s$}
\put(104,23){$\scriptstyle s$}
\put(83.5,-6.5){$\scriptstyle\bar s$}
\put(103.5,-6){$\scriptstyle s$}
\put(125,8){$=$}
\put(155,0){\line(1,1){20}}
\put(175,0){\line(-1,1){20}}
\put(165,10){\circle*{5}}
\put(154,23){$\scriptstyle s$}
\put(174,23){$\scriptstyle\bar s$}
\put(153.5,-6){$\scriptstyle s$}
\put(173.5,-6.5){$\scriptstyle\bar s$}
\end{picture}

\smallskip

\noindent
and also using the noncommuting symbols:
\bea
A_s(\theta_1)A_s(\theta_2)&=&
S(\theta)A_s(\theta_2)A_s(\theta_1)\nn\\
A_s(\theta_1)A_{\bar s}(\theta_2)&=&
S_T(\theta)A_{\bar s}(\theta_2)A_s(\theta_1)+
S_R(\theta)A_s(\theta_2)A_{\bar s}(\theta_1)~. \nn
\eea
Unitarity and crossing constrain these amplitudes. 
As a simple exercise, it is worthwhile to check that unitarity amounts
to
\bea
&& S(\theta)S(-\theta)=1\nn\\
&&S_T(\theta)S_T(-\theta)+S_R(\theta)S_R(-\theta)=1\nn\\
&&S_T(\theta)S_R(-\theta)+S_R(\theta)S_T(-\theta)=0\nn
\eea
while crossing is 
\bea
S(i\pi-\theta)\,&=&S_T(\theta)\nn\\
S_R(i\pi-\theta)&=&S_R(\theta)\nn
\eea
In 1977, Zamolodchikov was able to build on the earlier proposal of
Korepin and Faddeev (1975) for the special points $h=2n$,
$n\in\bbbn$ (at which $S_R(\theta)$ vanishes),
to conjecture an exact formula for the S-matrix. Subsequent
derivations made use of the Yang-Baxter equation, but in any event I
only want to quote the physical pole structure here. (A pole is called
`physical' if, like the crosses on figure~\ref{physst},
it lies on the physical strip.) A moment's thought
about the ways that the vacua fit together shows that:

\smallskip\noindent $\bullet$
$S_T$ can only form breathers in the forward channel;

\smallskip\noindent $\bullet$
$S$ can only form breathers in the crossed channel;

\smallskip\noindent $\bullet$
$S_R$ can form both.

\smallskip

This is precisely matched by Zamolodchikov's S-matrix: in terms of
$B(\beta)=2\beta^2/(8\pi{-}\beta^2)=4/h$, the poles of $S_T$, $S$ and
$S_R$ in the physical strip are found at the following points:

\smallskip

\noindent $\bullet$
$S_T$: $(1-k\frac{B}{2})\pi i$, $k=1,2,\dots\,$:

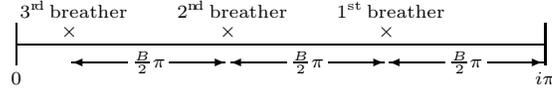
\begin{picture}(300,43)(-40,-23)
\thinlines
\put(80,0){\line(1,0){200}}
\put(80,-8){\line(0,1){16}}
\put(280,-8){\line(0,1){16}}
\put(80,-15){\makebox(0,0)[b]{$\scriptstyle 0$}}
\put(280,-15){\makebox(0,0)[b]{$\scriptstyle i\pi$}}
\put(100,3){\makebox(0,0)[b]{$\scriptstyle\times$}}
\put(100,10){\makebox(0,0)[b]{$\scriptstyle~3^{\rm r\!d}\,{\rm breather}$}}
\put(160,3){\makebox(0,0)[b]{$\scriptstyle\times$}}
\put(160,10){\makebox(0,0)[b]{$\scriptstyle~2^{\rm n\!d}\,{\rm breather}$}}
\put(220,3){\makebox(0,0)[b]{$\scriptstyle\times$}}
\put(220,10){\makebox(0,0)[b]{$\scriptstyle~1^{\rm st}\,{\rm breather}$}}
\multiput(122,-7)(60,0){3}{\vector(-1,0){21}}
\multiput(130,-7)(60,0){3}{\makebox(0,0){$\scriptstyle\frac{B}{2}\pi$}}
\multiput(138,-7)(60,0){3}{\vector(1,0){21}}
\end{picture}

\smallskip\noindent $\bullet$
$S$: $k\frac{B}{2}\pi i$, $k=1,2,\dots\,$:

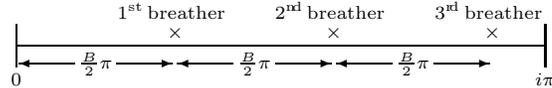
\begin{picture}(300,45)(-40,-27)
\thinlines
\put(80,0){\line(1,0){200}}
\put(80,-8){\line(0,1){16}}
\put(280,-8){\line(0,1){16}}
\put(80,-15){\makebox(0,0)[b]{$\scriptstyle 0$}}
\put(280,-15){\makebox(0,0)[b]{$\scriptstyle i\pi$}}
\put(140,3){\makebox(0,0)[b]{$\scriptstyle\times$}}
\put(140,10){\makebox(0,0)[b]{$\scriptstyle 1^{\rm st}\,{\rm breather}~$}}
\put(200,3){\makebox(0,0)[b]{$\scriptstyle\times$}}
\put(200,10){\makebox(0,0)[b]{$\scriptstyle 2^{\rm n\!d}\,{\rm breather}~$}}
\put(260,3){\makebox(0,0)[b]{$\scriptstyle\times$}}
\put(260,10){\makebox(0,0)[b]{$\scriptstyle 3^{\rm r\!d}\,{\rm breather}~$}}
\multiput(102,-7)(60,0){3}{\vector(-1,0){21}}
\multiput(110,-7)(60,0){3}{\makebox(0,0){$\scriptstyle\frac{B}{2}\pi$}}
\multiput(118,-7)(60,0){3}{\vector(1,0){21}}
\end{picture}

\smallskip\noindent $\bullet$
$S_R$: $(1{-}k\frac{B}{2})\pi i\,$, $k\frac{B}{2}\pi i$, $k=1,2,\dots\,$:

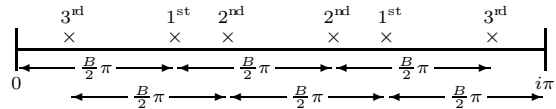
\begin{picture}(300,50)(-40,-30)
\thinlines
\put(80,0){\line(1,0){200}}
\put(80,-8){\line(0,1){16}}
\put(280,-8){\line(0,1){16}}
\put(80,-15){\makebox(0,0)[b]{$\scriptstyle 0$}}
\put(280,-15){\makebox(0,0)[b]{$\scriptstyle i\pi$}}
\put(140,3){\makebox(0,0)[b]{$\scriptstyle\times$}}
\put(140,10){\makebox(0,0)[b]{$\scriptstyle~1^{\rm st}$}}
\put(200,3){\makebox(0,0)[b]{$\scriptstyle\times$}}
\put(200,10){\makebox(0,0)[b]{$\scriptstyle~2^{\rm n\!d}$}}
\put(260,3){\makebox(0,0)[b]{$\scriptstyle\times$}}
\put(260,10){\makebox(0,0)[b]{$\scriptstyle~3^{\rm r\!d}$}}
\put(100,3){\makebox(0,0)[b]{$\scriptstyle\times$}}
\put(100,10){\makebox(0,0)[b]{$\scriptstyle~3^{\rm r\!d}$}}
\put(160,3){\makebox(0,0)[b]{$\scriptstyle\times$}}
\put(160,10){\makebox(0,0)[b]{$\scriptstyle~2^{\rm n\!d}$}}
\put(220,3){\makebox(0,0)[b]{$\scriptstyle\times$}}
\put(220,10){\makebox(0,0)[b]{$\scriptstyle~1^{\rm st}$}}
\multiput(102,-7)(60,0){3}{\vector(-1,0){21}}
\multiput(110,-7)(60,0){3}{\makebox(0,0){$\scriptstyle\frac{B}{2}\pi$}}
\multiput(118,-7)(60,0){3}{\vector(1,0){21}}
\multiput(122,-18)(60,0){3}{\vector(-1,0){21}}
\multiput(130,-18)(60,0){3}{\makebox(0,0){$\scriptstyle\frac{B}{2}\pi$}}
\multiput(138,-18)(60,0){3}{\vector(1,0){21}}
\end{picture}

\medskip

\noindent
(Beyond the physical strip, $S_T$, $S$ and $S_R$ have a proliferating
set of unphysical poles, there to fix up crossing and unitarity, but
this aspect will not be important below.) In the illustrations, the particles
responsible for the poles have also been indicated. To check that
these have been placed correctly, all that is needed is some
elementary kinematics. Suppose that a soliton $s$ and an antisoliton
$\bar s$, of masses $M_s=M_{\bar s}=M$ and moving
with respective rapidities $\theta_1$ and $\theta_2=-\theta_1$, 
fuse to form a (stationary) breather of mass $M_b$. 
The relative rapidity of the two particles is
$\theta_{12}=2\theta_1$, and the S-matrix will normally
have a simple, forward-channel pole at exactly this point. 
Conservation of energy dictates that
$M_b=2M\cosh(\theta_{12}/2)\,$. 
It will be
convenient to write this special value of $\theta_{12}$ as $iU_{s\bar
s}^b$,
where $U_{s\bar s}^b$ is called the fusing angle for the fusing $s\,\bar
s\rightarrow b$:

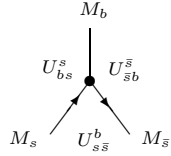
\begin{picture}(300, 75)(-70,5)
\thinlines
\put(85,40){\circle*{4}}
\put(60,16){\makebox(0,0)[b]{$\scriptstyle M_s$}}
\put(110,16){\makebox(0,0)[b]{$\scriptstyle M_{\bar s}$}}
\put(85,65){\makebox(0,0)[b]{$\scriptstyle~M_b$}}
\put(70,20){\line( 3, 4){ 14}}
\put(100,20){\line( -3,4){ 14}}
\put(81.25,35){\vector( 3, 4){0}}
\put(92.5,30){\vector( 3,-4){ 0}}
\put(85,40){\line( 0, 1){ 20}}
\put(85,22){\makebox(0,0)[t]{$\scriptstyle ~U_{s\bar s}^b$}}
\put(73,48){\makebox(0,0)[t]{$\scriptstyle U_{bs}^{s}$}}
\put(98,48){\makebox(0,0)[t]{$\scriptstyle U_{\bar sb}^{\bar s}$}}
\end{picture}

\noindent
By convention, an arrow pointing forwards in time marks 
a soliton, and an arrow pointing backwards
an antisoliton; lines without arrows are breathers of some sort.
Rotating the diagram by $\pm 2\pi/3$ gives pictures
of $b\,s$ and $\bar s\,b$ scattering, and the corresponding fusing
angles have also been indicated. If all of the
poles in $S_T$ are forward-channel, then the values of the fusing
angles follow from the positions of these poles:
\[
U_{s\bar s}^b=\left(1-k\frac{B}{2}\right)\pi\quad,\qquad
U_{bs}^{s}=U_{\bar sb}^{\bar s}=
\left(\frac{1}{2}+k\frac{B}{4}\right)\pi\,.
\]
The angles are all real, reflecting the fact that the bound
states are below threshold and the relative rapidities at which they
are formed purely imaginary.
The masses of the corresponding bound states are therefore
\[
M_b=2M\cos\left(\frac{\pi}{2}{-}k\frac{B}{4}\pi\right)=
2M\sin\left(\frac{k\pi}{h}\right)\,,
\]
and these match the spectrum of breather masses.

For later use, the precise relationship between $S_T$, $S$ and $S_R$
is:
\[
S_T(iu)=S(i\pi{-}iu)=\frac{\sin(\frac{2}{B}u)}{\sin(\frac{2}{B}\pi)}
S_R(iu)\,.
\]
The first equality is merely crossing symmetry, whilst the factor of
$\sin(\frac{2}{B}u)$ multiplying $S_R$ is there to
exclude the crossed-channel poles in $S_R$ from 
$S_T$, and the forward-channel poles in $S_R$ from $S$.

There are also S-matrix elements involving the breathers. These can be
deduced using bootstrap equations, to be described a little later, but
for now the focus is elsewhere and so I'll just quote the required 
result, concerning the scattering of two copies of the first breather:

\smallskip\noindent $\bullet$
$S_{11}$ has poles at $i\frac{B}{2}\pi$
and $i\pi-i\frac{B}{2}\pi$:

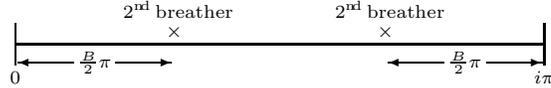
\nobreak
\begin{picture}(300,50)(-40,-20)
\thinlines
\put(80,0){\line(1,0){200}}
\put(80,-8){\line(0,1){16}}
\put(280,-8){\line(0,1){16}}
\put(80,-15){\makebox(0,0)[b]{$\scriptstyle 0$}}
\put(280,-15){\makebox(0,0)[b]{$\scriptstyle i\pi$}}
\put(140,3){\makebox(0,0)[b]{$\scriptstyle\times$}}
\put(140,10){\makebox(0,0)[b]{$\scriptstyle~2^{\rm n\!d}\,{\rm breather}$}}
\put(220,3){\makebox(0,0)[b]{$\scriptstyle\times$}}
\put(220,10){\makebox(0,0)[b]{$\scriptstyle~2^{\rm n\!d}\,{\rm breather}$}}
\multiput(102,-7)(140,0){2}{\vector(-1,0){21}}
\multiput(110,-7)(140,0){2}{\makebox(0,0){$\scriptstyle\frac{B}{2}\pi$}}
\multiput(118,-7)(140,0){2}{\vector(1,0){21}}
\end{picture}
\goodbreak

This looks fine: it is easy to check that the pole at $i\frac{B}{2}\pi$
can be blamed on a copy of the
second breather as a forward-channel bound state, and the
other one on the same particle appearing in the crossed channel. But
now consider what happens as $\beta$, and hence $B/2$, increases.
Each time $B/2$ passes an inverse integer, a pole in $S_T$ leaves the
physical strip and the corresponding breather leaves the spectrum of
the model. Finally, when $B/2$ passes $1/2$, the second breather
drops out. The theory, now well into the quantum regime, has just the
soliton, the antisoliton, and the first breather in its spectrum. And
this is problematical: $S_{11}$ still has a pair of
simple poles. How can this be, if the particle previously invoked to
explain them is no longer there?

The answer was found by Coleman and Thun in 1978, and requires a
preliminary diversion into the subject of anomalous
threshold singularities. These are most simply understood by asking
how an individual Feynman diagram might become singular. If the
external momenta are such that a number of internal propagators can
find themselves simultaneously on-shell, then it turns out that the
loop integrals give rise to a singularity in the amplitude. Apart from
the somewhat trivial examples provided by tree-level diagrams,
these singularities are always branch points
in spacetimes of dimension higher than two; but
in 1+1 dimensions, they can give rise to poles instead.

Once this is known, the problem of identifying the positions of such
singularities becomes a geometrical exercise in gluing together a
collection of on-shell vertices so as to make a pattern that closes.
For three-point vertices, the on-shell requirement simply forces the
relative Minkowski momenta to be equal to $i$ times the fusing angles. 
If all couplings are below threshold, then all fusing angles are real
and the resulting patterns can be drawn as figures in two Euclidean 
dimensions. These pictures are known as Landau, or on-shell, diagrams.

In fact, the characterisation as so far given also encompasses the more
usual multiparticle thresholds, which are associated in perturbation
theory with on-shell diagrams of the
following type:

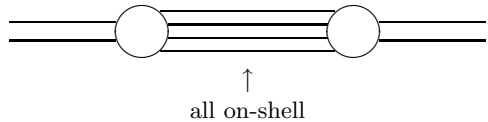
\begin{picture}(300,65)(-10,-38)
\thinlines
\multiput(70,-1)(0,7){2}{\line(1,0){40}}
\multiput(130,0)(0,5){2}{\line(1,0){60}}
\multiput(127,-5)(0,15){2}{\line(1,0){66}}
\multiput(210,-1)(0,7){2}{\line(1,0){40}}
\put(120,2.5){\circle{21}}
\put(200,2.5){\circle{21}}
\put(160,-16){\makebox(0,0){$\uparrow$}}
\put(160,-27){\makebox(0,0){\small all on-shell}}
\end{picture}

\noindent
(Exceptionally, time is running sideways in this picture.) Here, 
the on-shell particles are all in the same channel, and the value of
$s$ at which the singularity is found is simply the square of the sum
of the masses of the intermediate on-shell particles. To qualify as
`anomalous', something more exotic should be going on, and the position
of the singularity will no longer have such a straightforward relationship 
with the mass spectrum of the model.

The moral is that when we come to analyse the pole
structure of an S-matrix in 1+1 dimensions there are more things to
worry about than just the tree-level processes discussed so far.
Returning to the sine-Gordon model, as the point $B/2=1/2$ is passed,
an on-shell diagram does indeed enter the game as far as the
scattering of two of the first breathers is concerned:

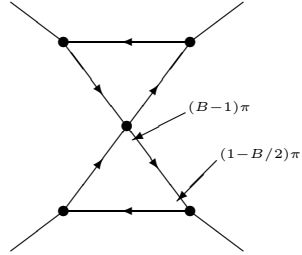
\begin{picture}(300,110)(-60,-20)
\thinlines
\put(60,0){\line(1,0){48}}
\put(60,0){\line(3,4){48}}
\put(75,20){\vector(3,4){0}}
\put(96,48){\vector(3,4){0}}
\put(108,0){\line(-3,4){48}}
\put(75,44){\vector(3,-4){0}}
\put(96,16){\vector(3,-4){0}}
\put(60,64){\line(1,0){48}}
\put(40,-15){\line(4,3){20}}
\put(108,64){\line(4,3){20}}
\put(108,0){\line(4,-3){20}}
\put(40,79){\line(4,-3){20}}
\put(82,0){\vector(-1,0){0}}
\put(82,64){\vector(-1,0){0}}
\put(118,19){\vector(-1,-1){15}}
\put(119,20){\makebox(0,0)[bl]{$\scriptscriptstyle (1{-}B/2)\pi$}}
\put(106,37){\vector(-2,-1){20}}
\put(107,38){\makebox(0,0)[bl]{$\scriptscriptstyle (B{-}1)\pi$}}
\put(60,0){\circle*{4}}
\put(60,64){\circle*{4}}
\put(108,64){\circle*{4}}
\put(108,0){\circle*{4}}
\put(84,32){\circle*{4}}
\end{picture}

\noindent
This diagram only invokes the solitons and antisolitons on the
internal lines, which are present in the spectrum whatever the
coupling.
A couple of internal angles are marked,
making it clear that the figure will only close if $B/2\ge 1/2$. 
However we are not quite out of the woods yet: diagrams of this sort
are expected to yield double poles when evaluated in 1+1 dimensions, 
and
not the single poles that we are after as soon as $B/2$ passes $1/2$.
(Actually at $B/2=1/2$, $S_{11}$ does indeed have a double pole, but
the understanding of a single extra point is scarcely major progress.)
The final ingredient is to notice that for
$B/2>1/2$, two of the internal lines must inevitably cross over. 
When a soliton and an antisoliton meet we should
allow for reflection as well as transmission, since we have already
seen that both amplitudes are generally nonzero. Thus not one but four
diagrams are relevant to the
amplitude near to the value of $\theta_{12}$ of interest. The full
story is given by the diagrams

{\setlength{\unitlength}{0.6pt}
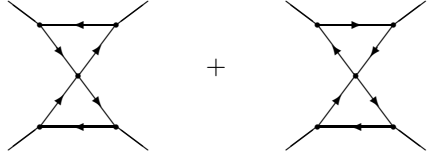
\begin{picture}(300,115)(-30,-30)
\put(0,0){
\begin{picture}(300,100)(-60,5)
\thinlines
\put(60,0){\line(1,0){48}}
\put(60,0){\line(3,4){48}}
\put(75,20){\vector(3,4){0}}
\put(99,52){\vector(3,4){0}}
\put(108,0){\line(-3,4){48}}
\put(75,44){\vector(3,-4){0}}
\put(99,12){\vector(3,-4){0}}
\put(60,64){\line(1,0){48}}
\put(40,-15){\line(4,3){20}}
\put(108,64){\line(4,3){20}}
\put(108,0){\line(4,-3){20}}
\put(40,79){\line(4,-3){20}}
\put(81,0){\vector(-1,0){0}}
\put(81,64){\vector(-1,0){0}}
\put(60,0){\circle*{4}}
\put(60,64){\circle*{4}}
\put(108,64){\circle*{4}}
\put(108,0){\circle*{4}}
\put(84,32){\circle*{4}}
\end{picture}}
\put(230,28){$+$}
\put(175,0){
\begin{picture}(300,100)(-60,5)
\thinlines
\put(60,0){\line(1,0){48}}
\put(60,0){\line(3,4){48}}
\put(75,20){\vector(3,4){0}}
\put(93,44){\vector(-3,-4){0}}
\put(108,0){\line(-3,4){48}}
\put(69,52){\vector(-3,4){0}}
\put(99,12){\vector(3,-4){0}}
\put(60,64){\line(1,0){48}}
\put(40,-15){\line(4,3){20}}
\put(108,64){\line(4,3){20}}
\put(108,0){\line(4,-3){20}}
\put(40,79){\line(4,-3){20}}
\put(81,0){\vector(-1,0){0}}
\put(88,64){\vector(1,0){0}}
\put(60,0){\circle*{4}}
\put(60,64){\circle*{4}}
\put(108,64){\circle*{4}}
\put(108,0){\circle*{4}}
\put(84,32){\circle*{4}}
\end{picture}}
\end{picture}
}

\noindent
together with their overall conjugates, in which all arrows are
reversed. Individually each diagram 
contributes a double pole, but these must be added together
with the correct relative weights. The only difference between the two
diagrams shown is that the central blob on the first carries with it a
factor of $S_T(\theta)$, and the central blob of the
second a factor of $S_R(\theta)$.
At the pole position,
the value of $\theta$ is fixed by the
on-shell requirement to be equal to $iu$, with $u=(B{-}1)\pi$. 
Referring back to the earlier expression relating $S_T$ to $S_R$,
we have
\bea
\openup 1\jot
S_T((B{-}1)\pi i)&=&
\frac{\sin(\frac{2}{B}(B{-}1)\pi)}{\sin(\frac{2}{B}\pi)}
S_R((B{-}1)\pi i)\nn\\
&=&- S_R((B{-}1)\pi i)\nn
\eea
Thus exactly when the individual diagrams have a double pole, 
$S_T+S_R=0$, a cancellation occurs, and the 
field-theoretic prediction is for a simple pole, exactly as seen in
the S-matrix. Coleman and Thun dubbed explanations of this sort
`prosaic', since they do not rely on properties special to
integrable field theories -- a 
non-integrable (albeit very finely-tuned)
theory would be perfectly capable of exhibiting the same behaviour. 
Nonetheless, there is a certain miraculous quality about the result.
The cancellation between $S_T$ and $S_R$ is very delicate: $S_T$ 
describes a classically-allowed process,
while $S_R$ does not (there is no classical
reflection of solitons). It is also noteworthy that the
Landau diagrams expose intrinsically field-theoretical aspects of the
theory, since loops are involved. Their relevance 
tells us that quantum mechanical intuitions about bound
states and pole structure may occasionally be misleading.

\noindent
Some general lessons can be drawn from all of this:

\smallskip\noindent $\bullet$
The S-matrix can have poles between $\theta=0$ and $\theta=i\pi\,$;

\smallskip\noindent $\bullet$
these can be first order, second order, or in fact much higher order
(examples up to $12^{\rm th}$ order are found in the affine Toda field
theories);

\smallskip\noindent $\bullet$
even for a first-order pole, a direct interpretation in terms of a
bound state is not inevitable;

\smallskip\noindent $\bullet$
but there is always some (prosaic$^{\rm TM}$) explanation in terms of
standard field theory.

\section{Bootstrap equations}
If we decide that our theory does contain a bound state, then 
the next task is to find the S-matrix elements involving this new
particle, and then to
look for evidence of further bound states in these, and so
on. Rather than continuing with the sine-Gordon example, which showed
how complicated the story can become, I will make a tactical
retreat at this point to a class of models where the behaviour is
rather simpler, and the workings of the bound states can be seen more
cleanly. The structure is still rewardingly rich, so this won't be
too great a sacrifice. 

The key concession is to assume that there are no
degeneracies among the one-particle states once all of the non-zero
spin conserved charges have been specified. This closes off the
loophole exploited by the sine-Gordon model, and forces the scattering
to be diagonal. 

The S-matrix now only needs two indices:

{\setlength{\unitlength}{0.7pt}
\begin{picture}(300,110)(-150,-55)
\thinlines
\put(-40,-6){$S_{ij}(\theta_1-\theta_2)~~~~=$}
\put(90,-32){\vector(3,4){22}}
\put(138,-32){\vector(-3,4){22}}
\put(114,0){\vector(3,4){23}}
\put(114,0){\vector(-3,4){23}}
\put(114,0){\circle*{6}}
\put(90,-36){\makebox(0,0)[t]{$\scriptstyle A_i(\theta_1)$}}
\put(138,-36){\makebox(0,0)[t]{$\scriptstyle A_j(\theta_2)$}}
\put(90,36){\makebox(0,0)[b]{$\scriptstyle A_j(\theta_2)$}}
\put(138,36){\makebox(0,0)[b]{$\scriptstyle A_i(\theta_1)$}}
\end{picture}}

Two of the previous constraints on the two-particle S-matrix 
elements can therefore be simplified:

\noindent $\bullet$ Unitarity:\quad $S_{ij}(\theta)S_{ij}(-\theta)=
1\,;$
\smallskip

\noindent $\bullet$ Crossing:\quad\, $S_{ij}(\theta)=
S_{i\bar\jmath}(i\pi-\theta)\,.$
\smallskip

\noindent
(In contrast to its previous incarnation, there is no sum on repeated
indices in the unitarity equation.) Combining these two reveals the
important fact that
\[
S_{ij}(\theta{+}2\pi i)=S_{ij}(\theta)\,,
\]
so that for diagonal scattering the Riemann surface for the S-matrix
really is just a double cover of the complex plane -- whether you go
round the left or the right branch point in figure~\ref{physs}, you
always land up on the same unphysical sheet.

The simplification is even more drastic for the third
constraint:
the loss of matrix structure, already evident in the revised
unitarity equation, means that the Yang-Baxter equation
is trivially satisfied for any
$S_{ij}(\theta)$ whatsoever.

Fortunately, a vestige of 
algebraic structure does remain, in the guise of the
pattern of bound states. Suppose that $S_{ij}(\theta_{12})$ has a simple
pole, at
$\theta_{12}=iU^k_{ij}$ say, which really is due to the formation of a
forward-channel bound state. Note that, in a unitary theory, forward
and crossed channel poles can be distinguished by the fact that the
residues are positive-real multiples of $i$ in the forward channel,
and negative-real multiples in the crossed channel.
The previous picture of the
scattering process can be `expanded' near to the pole:

{\setlength{\unitlength}{0.7pt}
\begin{picture}(300,165)(-160,-55)
\thinlines
\put(-48,16){$S_{ij}(\theta_{12}\approx iU^k_{ij})~~~~\sim$}
\put(90,-32){\vector(3,4){22}}
\put(138,-32){\vector(-3,4){22}}
\put(114,34){\vector(3,4){23}}
\put(114,34){\vector(-3,4){23}}
\put(114,0){\vector(0,1){22}}
\put(114,0){\line(0,1){34}}
\put(114,0){\circle*{6}}
\put(114,34){\circle*{6}}
\put(90,-36){\makebox(0,0)[t]{$\scriptstyle A_i(\theta_1)~$}}
\put(138,-36){\makebox(0,0)[t]{$\scriptstyle ~A_j(\theta_2)$}}
\put(90,72){\makebox(0,0)[b]{$\scriptstyle A_j(\theta_2)~$}}
\put(138,72){\makebox(0,0)[b]{$\scriptstyle ~A_i(\theta_1)$}}
\put(122,18){\makebox(0,0)[l]{$\scriptstyle ~A_{\bar k}(\theta_3)$}}
\end{picture}}

\noindent
(The intermediate particle is labelled $\bar k$ for convenience,
in anticipation of a convention that all indices on a three-point 
coupling will be ingoing.)
There are a number of immediate consequences:
\smallskip

\noindent $(1)$ The quantum coupling $C^{ijk}$ is nonzero at the point
where particles $i$, $j$ and $k$ are all on shell.
\smallskip

\noindent $(2)$ At the rapidity difference $\theta_{12}=iU^k_{ij}\,$,
the intermediate particle $A_{\bar k}(\theta_3)$ is on shell and survives 
for macroscopic times. On general grounds (the `bootstrap principle', or
`nuclear democracy'), $A_{\bar k}$ is expected to be one of the other
asymptotic one-particle states of the model.
\smallskip

\noindent $(3)$ Since $s=m_k^2$ when this happens, we have
\[
m_k^2=m_i^2+m_j^2+2m_im_j\cos U^k_{ij}\,.
\]
Of course, the $U_{ij}^k$ are just the fusing angles already seen in
the last lecture.
\smallskip

\noindent
For a more geometrical characterisation of the fusing angles,
observe that the formula just given is familiar from elementary
trigonometry, and implies that $U^k_{ij}$ is the outside angle of a
`mass triangle' of sides $m_i$, $m_j$ and $m_k\,$:

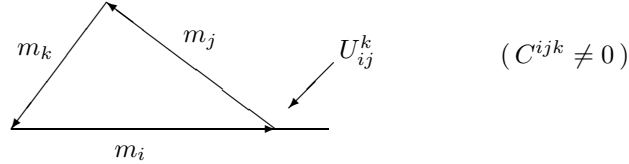
\begin{picture}(300,75)(-25,-15)
\thinlines
\put(20,0){\vector(1,0){100}}
\put(120,0){\line(1,0){20}}
\put(120,0){\vector(-4,3){64}}
\put(56,48){\vector(-3,-4){36}}
\put(65,-6){\makebox(0,0)[t]{$m_i$}}
\put(85,33){\makebox(0,0)[bl]{$m_j$}}
\put(36,28){\makebox(0,0)[br]{$m_k$}}
\put(142,25){\vector(-1,-1){17}}
\put(144,27){\makebox(0,0)[bl]{$U^k_{ij}$}}
\put(205,28){\makebox(0,0)[l]{$(\,C^{ijk}\neq 0\,)$}}
\end{picture}

\noindent
With $C^{ijk}\neq 0\,$, poles are also present in $S_{jk}$ and
$S_{ki}\,$. From the triangle just drawn, the three fusing angles
involved satisfy
\[
U^k_{ij}+U^i_{jk}+U^j_{ki}=2\pi~.
\]
Concrete examples, the nonzero quantum couplings $C^{bs\bar s}$ in the
sine-Gordon model, were mentioned in the last lecture.

Since the $\bar k$ is supposed to be long-lived when
$\theta_{12}=iU^k_{jk}$, it should be possible to evaluate a
conserved charge $Q_s$ after the fusing of $i$ and $j$ into $\bar k$,
as well as before. The action of $Q_s$ on 
$\ket{A_{\bar k}(\theta_3)}$ 
and 
$\ket{A_i(\theta_1)A_j(\theta_2)}$ was given at the beginning of
the second lecture; equating the two at the relevant rapidities gives
a constraint on the numbers $q^{(s)}_i$ which characterise $Q_s$:
\[
\quad C^{ijk}\neq 0~~\Rightarrow\qquad
q^{(s)}_{\bar k}=q^{(s)}_ie^{is\bar U^j_{ki}}
+q^{(s)}_je^{-is\bar U^i_{kj}},\quad~~\qquad
\]
where $\bar U=\pi{-}U$. (To see this, switch to the frame where $\bar
k$ is stationary. Then $\theta_1=i\bar U^j_{ki}$ and 
$\theta_2=-i\bar U^i_{kj}$.) The relations $q^{(s)}_{\bar
k}=(-1)^{s+1}q^{(s)}_k$ and $\bar U+\bar U+\bar U=\pi$ can be employed
to put this into a more symmetrical form:
\[
\quad C^{ijk}\neq 0~~\Rightarrow\qquad
q^{(s)}_i+q^{(s)}_je^{isU^k_{ij}}
+q^{(s)}_ke^{is(U^k_{ij}+U^i_{jk})}=0\,.
\]
Drawing this equation in the complex plane shows that it has a nice
interpretation as a closure condition for a `generalised mass 
triangle':

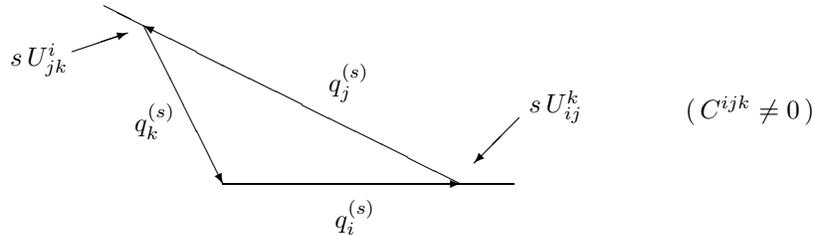
\begin{picture}(300,110)(-55,-25)
\thinlines
\put(30,0){\vector(1,0){90}}
\put(120,0){\line(1,0){20}}
\put(120,0){\vector(-2,1){120}}
\put(0,60){\vector(1,-2){30}}
\put(0,60){\line(-2,1){15}}
\put(80,-6){\makebox(0,0)[t]{$q^{(s)}_i$}}
\put(70,35){\makebox(0,0)[bl]{$q^{(s)}_j$}}
\put(12,20){\makebox(0,0)[br]{$q^{(s)}_k$}}
\put(142,25){\vector(-1,-1){17}}
\put(144,27){\makebox(0,0)[bl]{$\,s\,U^k_{ij}$}}
\put(-27,50){\vector(3,1){21}}
\put(-27,47){\makebox(0,0)[r]{$s\,U^i_{jk}\,$}}
\put(205,28){\makebox(0,0)[l]{$(\,C^{ijk}\neq 0\,)$}}
\end{picture}

\noindent
(Note that the three angles for this triangle,
$s\,U^k_{ij}$, $s\,U^i_{jk}$ and $s\,U^j_{ki}\,$, now add up to $2\pi s$
instead of just $2\pi$.)

This set of equations constitutes the conserved charge bootstrap.
Given a set of masses and three-point couplings, the fusing angles
can be determined by the mass triangles. The angles in the higher-spin
triangles are then fixed, and at any given spin $s$ the demand that
all of the triangles at that spin should
close provides an overdetermined set of conditions on the values of
the $q^{(s)}_i$. Should the only solution be the trivial one,
$q^{(s)}_i=0$ for all $i$, then we can conclude that $Q_s\equiv 0$ and
there is no conserved charge of that spin.
The surprise is that there should be {\it any} choice
of the initial masses and couplings such that
the higher-spin triangles can be made
to close for at least an infinite subset of spins.
However, when this does happen,
the set of spins at which the triangles do
close gives access to the fingerprint of spins
mentioned earlier, without the need to find the
local conserved densities explicitly.

A fair amount of physical intuition has been used to arrive at these
conclusions, and one particular point deserves mention. Given that the
fusing angles are always real in applications of interest,
most if not all of the momenta entering the discussion are complex,
and this might cast doubt on the spacetime language
that has been used throughout.
I have implicitly assumed that the states, their
fusings, and the
action on them of the conserved charges, continue to behave in the
expected manner
after the necessary analytic continuations have been made.

For the S-matrix we can use a similar argument. Consider another
particle $l$, which might interact either before or after particles
$i$ and $j$ fuse to form $\bar k$. Which depends on the impact
parameter, but in an integrable model this should be irrelevant.
Translating this into equations,
\[
C^{ijk}\neq 0~~\Rightarrow\qquad
S_{l\bar k}(\theta)=
S_{li}(\theta-i\bar U^j_{ki})S_{lj}(\theta+\bar U^i_{jk})\,,\qquad\quad
\]
and this is the S-matrix bootstrap equation. It
can be given a more symmetrical appearance using crossing
symmetry and unitarity, becoming:
\[
\quad C^{ijk}\neq 0~~\Rightarrow\qquad
S_{li}(\theta)S_{lj}(\theta+iU^k_{ij})
S_{lk}(\theta+i(U^k_{ij}{+}U^i_{jk}))=1\,.\quad
\]
Imposing these relations for each non-vanishing three-point coupling
provides an overdetermined set of functional equations, and again it
is rather surprising that there are any solutions. In fact the two
bootstraps are rather directly related:

%\smallskip\noindent
%Exercise: 
\exercise{ with the help of a logarithmic derivative and a Fourier
expansion, show that each solution to the S-matrix bootstrap contains
within it a solution to the corresponding conserved charge bootstrap.}
%\smallskip

Starting from an initial guess of a single S-matrix element, we can
now search for poles, infer some three-point couplings, apply the
bootstrap to deduce further S-matrix elements, and then iterate 
away. If the process closes on a finite set of particles, then we can
chalk up a success and go on to another problem; if not, then the
initial guess should probably be revised. This is precisely the
approach that A.B.Zamolodchikov took in his pioneering work, 
(1989a) and (1989b), on perturbed conformal field theories.
The next lecture is devoted to a particularly interesting example of
this procedure which relates to the behaviour of the $T=T_c$ Ising model,
in a small magnetic field.

\section{Zamolodchikov's $E_8$-related S-matrix}
The critical Ising model is found at zero magnetic field, with the
temperature carefully adjusted to the critical value $T_c$. If the
continuum limit is taken at this point, the result is well-known to be
described by the $c=\half$ conformal field theory, a very
well-understood object. In the papers (1989a) and (1989b), Zamolodchikov
probed nearby points by considering the
actions
\[
S_{\rm pert}=S_{\rm CFT}+\lambda\!\int\!d^2x\phi(x)\,,
\]
where $S_{\rm CFT}$ is a notional action for the $c=\half$ conformal
field theory, inside of which $\phi$ sits as one of the spinless,
relevant fields. There are just two of these for the Ising model, and
one of them, usually labelled $\sigma$, can be identified with the
scaling limit of the local magnetisations (spins) on the lattice. 
Thus perturbing by $\sigma$ corresponds to switching on a magnetic
field. The game now is to exploit the great control that we have of
the unperturbed situation to divine some information about the
perturbed model. In this particular case, Zamolodchikov used an
ingenious argument, based on the counting of dimensions in Virasoro
representations, to establish that the perturbed model supported
at the very least
local conserved charges with the following spins:
\[
s~~=~~1,~7,~11,~13,~17,~19.
\]
This tells us two things. First, there are certainly enough
charges here to employ Parke's argument, and so the perturbed model,
if massive, possesses a factorisable S-matrix (and the model must be
massive, since the $c$-theorem tells us that the central charge of any
conformal infrared limit would be less than $\half$, and there is no
such unitary conformal field theory).
Second, we now have the first part of the fingerprint of conserved
spins, and can hope to use this information to build a bridge between
the ultraviolet information residing in the characterisation of the
model as a perturbed conformal field theory, and the infrared
information that would be revealed if we knew its S-matrix.

To commence the search for this S-matrix, suppose that the massive
theory possesses a particle of mass $m_1$, say. In addition, assume
for the time being that the model falls into the simplest class, that
of diagonal scattering theories. The magnetic field breaks the
$\bbbz_2$ symmetry of the unperturbed model, and so there is no reason
to exclude an interaction of $\phi^3$ type from the effective
Lagrangian of the perturbed theory. This places the model in the same
general class as the second example discussed in the first lecture, 
and makes it natural for
$C^{111}$ to be nonzero. For this coupling the mass triangle is
equilateral, and the fusing angles are therefore
all equal to $2\pi/3$. The
conserved charge bootstrap equation is
\[
C^{111}\neq 0~~\Rightarrow\qquad
q^{(s)}_1+q^{(s)}_1e^{2\pi is/3}
+q^{(s)}_1e^{4\pi is/3}=0\,.
\]
This equation has a nontrivial solution whenever $s$ has no common
divisor with~$6\,$:
\[
s~~=~~1,~5,~7,~11,~13,~17\dots~.
\]
This is too much of a good thing: the fingerprint
contains rather too many spins for comfort.
Whilst the unwanted charges might vanish for other reasons,
it would be more satisfying if the cast of particles could be enlarged
a little, so as to restrict the set of conserved spins a bit more.
Besides, earlier work described in McCoy and Wu (1978) had led
Zamolodchikov to suspect the presence of at least a couple of further
masses in the particle spectrum. Taking things one step at a time, he
first enlarged the spectrum by adding just one more particle type, with
mass $m_2$, and supposed that both $C^{112}$ and $C^{221}$ were nonzero.
The fusing angles are not so easily determined now, but if the
ignorance is encoded in the pair of numbers $y_1=\exp(iU^1_{21})$ and
$y_2=\exp(iU^2_{12})$, then two of the bootstrap equations are:
\[
C^{121}\neq 0~~\Rightarrow\qquad
q^{(s)}_1+q^{(s)}_2(y_1)^s +q^{(s)}_1(y_1)^{2s}=0\,;
\]
\[
C^{212}\neq 0~~\Rightarrow\qquad
q^{(s)}_2+q^{(s)}_1(y_2)^s +q^{(s)}_2(y_2)^{2s}=0\,.
\]
Eliminating $q^{(s)}_1$ and $q^{(s)}_2$,
\[
(y_1^s+y_1^{-s}) (y_2^s+y_2^{-s})=1\,,
\]
at least at those values of $s$ for which there is a nontrivial
conserved charge. If there are more than a couple of these, then
the system is overdetermined; nevertheless, if
$y_1=\exp(4\pi i/5)$ and $y_2=\exp(3\pi i/5)$ then there is a solution
for every odd $s$ which is not a multiple of~$5$. This yields the
following set of fusing angles:
\[
U^1_{12}=U^1_{21}=4\pi/5~,\quad U^2_{11}=2\pi/5~;\nn\\
\]
\[
U^2_{21}=U^2_{12}=3\pi/5~,\quad U^1_{22}=4\pi/5~,\nn
\]
and the golden mass ratio
\[
\frac{m_2}{m_1}=2\cos\frac{\pi}{5}~.
\]
(There are other solutions such as $(-y_1,-y_2)$ or $(y_2,y_1)$,
but the choice taken is the
only one which yields sensible fusing angles and $m_1<m_2$.)

This is very promising: the multiples of $5$ were exactly the values
of $s$ that had to be eliminated in order to match the sets of
conserved spins. Thus
encouraged,  we can start to think about the S-matrix. 

It has been assumed that all the particles are self-conjugate (the
absence of any even spins from the fingerprint is a good hint that
this assumption is correct) and so each S-matrix element $S_{ij}$ must 
be individually crossing-symmetric,
$S_{ij}(\theta)=S_{ij}(i\pi{-}\theta)$, as well as unitary. It is
convenient to construct these as products of a basic `building block'
$(x)(\theta)$, where 
\[
(x)(\theta)=
\frac{\sinh\left(\frac{\theta}{2}+\frac{i\pi x}{60}\right)}
     {\sinh\left(\frac{\theta}{2}-\frac{i\pi x}{60}\right)}~.
\]
(The $60$ in the denominators has been chosen with advance knowledge
of the final answer, so that all of the arguments $x$ will turn out to be
integers.) Unitarity is built into these blocks, whilst the crossing 
symmetry just mentioned is assured if each block
$(x)$ is always accompanied by $(30{-}x)$. The block $(x)$ has
a single physical-strip pole at $i\pi x/30$, and no physical-strip
zeroes.

Consider first $S_{11}(\theta)$. The nonzero couplings $C^{111}$ and
$C^{112}$ imply forward-channel poles at $iU^1_{11}=2\pi i/3$ and
$iU^2_{11}=2\pi i/5$.
Incorporating these and their crossed partners
into a first guess for the S-matrix
element gives
\[
S_{11}=(10)(12)(18)(20)~.
\]
However this can't be the whole story. The S-matrix bootstrap equation
for the $\phi^3$ coupling $C^{111}$ requires that
$S_{11}(\theta{-}i\pi/3)S_{11}(\theta{+}i\pi/3)$ should be equal to 
$S_{11}(\theta)$. But it is easy to check that for the guess just
given,
\[
S_{11}(\theta{-}i\pi/3)S_{11}(\theta{+}i\pi/3)
=-(2)(8)(10)(20)(22)(28)~,
\]
which is not the desired answer. Stare at the equations long enough,
though, and you might just spot that all will be well if the initial
guess is multiplied by the factor $-(2)(28)$. Thus the
minimal solution to the constraints imposed so far is
\[
S_{11}=-(2)(10)(12)(18)(20)(28)~.
\]
The bootstrap equations have forced the addition of two extra poles, and
the simplest option is to suppose that these are the forward- and
crossed-channel signals of a further
particle, with mass $m_3=2m_1\cos(\pi/30)$, and a nonzero
coupling $C^{113}$. Of course, the story is not over yet. Using the
bootstrap for the fusing $1\,1\rightarrow 2$ allows $S_{12}$ to be
obtained from the provisional $S_{11}$:
\[
S_{12}=(6)(8)(12)(14)(16)(18)(22)(24)\,.
\]
The poles from the blocks $(14)$, $(18)$ and $(24)$ are
correctly-placed to match forward-channel copies of particles the 
$3$, $2$ and $1$ respectively, those in $(6)$, $(12)$ and $(16)$ can
then be blamed on the same particles in the crossed channel, but the
blocks $(8)$ and $(22)$ are not so easily dismissed, and require the
addition of yet another particle, of mass $m_4$ say. (Consideration of
the signs of the residues shows that the forward-channel pole is at
$8\pi i/30$.) Next, the bootstrap for $1\,1\rightarrow 3$ predicts
\[
S_{13}=(1)(3)(9)(11)^2(13)(17)(19)^2(21)(27)(29)\,.
\]
%(Exercise: 
\bexercise{ check at least one of these claims.}

\noindent
Apart from the double poles, which should not be too alarming after
the earlier investigations of the sine-Gordon model, there is one more
pair of simple poles here which cannot be explained in terms of the 
spectrum seen so far, and so a further mass, $m_5$ say, is revealed.

There is nothing to stop the mythical energetic reader from continuing
with all this, and it turns out that no further backtracking is required
-- with just one correction to the initial guess, Zamolodchikov had
arrived at a consistent conjecture for $S_{11}(\theta)$. Furthermore,
the final answer turned out to have
a number of intriguing properties. These can be
summarised in a list of what might be called `S-matrix data':

\smallskip\noindent
$\bullet$ $8$ particle types $A_1,\dots A_8$;

\smallskip\noindent
$\bullet$ $8$ masses $m_i$, $i=1,\dots 8$, which together form an
eigenvector of the Cartan matrix of the Lie algebra $E_8$:
\[
C^{[E_8]}_{ij}m_j=(2{-}2\cos\frac{\pi}{30})m_i~;
\]
(This allows each particle type to be attached to a spot on the $E_8$
Dynkin diagram -- more on this later.)

\smallskip\noindent
$\bullet$ solutions to the conserved-charge bootstrap found at
\[
s~~=~~1,~7,~11,~13,~17,~19,~23,~29~\dots
\]
thus fitting the fingerprint found from perturbed conformal field
theory (and also the exponents of $E_8$, repeated modulo $30$);

\smallskip\noindent
$\bullet$ `charges' associated with these solutions which form further
eigenvectors of $C^{[E_8]}_{ij}\,$:
\[
C^{[E_8]}_{ij}q^{(s)}_j=(2{-}2\cos\frac{\pi s}{30})q^{(s)}_i~;
\]

\smallskip\noindent
$\bullet$ a full two-particle S-matrix which is a collection of
complicated but elementary functions, with poles at integer multiples
of $i\pi/30$, all products of the
elementary building blocks introduced earlier. 

\smallskip
There is no space to record the full S-matrix here, but a complete
table can be found in, for example, Braden et al.~(1990).
\smallskip

One note of caution: elegant though it might be,
it is not completely clear that this is the answer to 
the question originally posed, given
the number of assumptions that were made along the way.
Probably the most convincing
reassurance comes on recalculating the central charge of the
unperturbed model from the conjectured
S-matrix, using a technique called the
thermodynamic Bethe ansatz. Its use in this context was first
advocated by Al.B.Zamolodchikov (1990), and the specific calculation for
the $E_8$-related S-matrix can be found in Klassen and Melzer (1990).

\section{Coxeter geometry}
It is a finite though lengthy task to check all of the bootstrap
equations for Zamolodchikov's S-matrix, and to verify the 
properties listed at the end of the last lecture. However there is
something not completely satisfactory about this, and a 
feeling that an underlying structure remains to be discovered, a
structure that might help to explain quite why such an elegant
solution to the bootstrap should exist at all. 
The purpose of this short section, something of an aside from the main
development, is to show that at least some parts of this
question can be answered.

One mathematical preliminary is required, a quick recap on the Weyl
group of $E_8$. Imagine a hedgehog $\Phi$ of $240$ vectors, or `roots',
sitting in eight dimensions. They all have equal length, and together
they make up the root system of $E_8$. Each root can be written as an
integer combination of the simple roots $\{\alpha_1,\dots\alpha_8\}$:
\[
\alpha\in\Phi\quad\Rightarrow\quad
\alpha=\sum_{i=1}^8m_i\alpha_i
\]
with $m_i\in\bbbz$, and the $m_i$ either all non-negative, or all
non-positive. Actually, there are $240$ different eight-element subsets
of $\Phi$ which could serve as the simple roots, but their geometrical
properties are all identical, and can be summarised by giving the set
of their mutual inner products, as encoded either in the Cartan matrix
\[
C^{[E_8]}_{ij}=2\frac{\alpha_i.\alpha_j}{~\alpha_j^2}~,
\]
or the Dynkin diagram

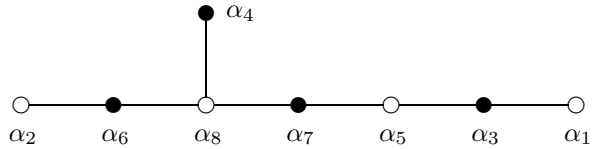
\begin{picture}(300,80)(-20,-25)
\thinlines
\multiput(20,0)(70,0){4}{\circle{6}}
\multiput(55,0)(70,0){3}{\circle*{6}}
\put(90,35){\circle*{6}}
\put(90,3){\line(0,1){29}}
\multiput(23,0)(35,0){6}{\line(1,0){29}}
\put(20,-9){\makebox(0,0)[t]{$\,\alpha_2$}}
\put(55,-9){\makebox(0,0)[t]{$\,\alpha_6$}}
\put(90,-9){\makebox(0,0)[t]{$\,\alpha_8$}}
\put(125,-9){\makebox(0,0)[t]{$\,\alpha_7$}}
\put(160,-9){\makebox(0,0)[t]{$\,\alpha_5$}}
\put(195,-9){\makebox(0,0)[t]{$\,\alpha_3$}}
\put(230,-9){\makebox(0,0)[t]{$\,\alpha_1$}}
\put(96,35){\makebox(0,0)[l]{$\,\alpha_4$}}
\end{picture}

\noindent
Pairs of simple roots joined by a line have inner product $-1$, and
all other pairs are orthogonal. In particular
this means that the black-coloured
roots are mutually orthogonal, as are the white-coloured roots.
The labelling might look random, but recall from the last lecture that
the vector of masses formed an eigenvector of the Cartan matrix, so
that each particle type in Zamolodchikov's S-matrix
can be assigned to a point on the Dynkin diagram. The
labels used here correspond to these particle labels, with
$m_1<m_2<\dots<m_8$.

For each $\alpha\in\Phi$, define the Weyl reflection
$r_{\alpha}$ to be the reflection in the $7$-dimensional
hyperplane orthogonal to $\alpha\,$:
\[
r_{\alpha}\,:\quad x\mapsto x-2\frac{\alpha.x}{~\alpha^2}\alpha\,.
\]
The products of the Weyl reflections in any order and of any length
together make up $W$, the Weyl group of $E_8$.
This group maps $\Phi$ to itself, and is
{\it finite}:
$|W|<\infty$. (Note that $W$ is therefore a finite reflection group,
a much simpler object than the Lie group or algebra with which it
is associated.)  One more fact: to generate $W$, it's enough
to start with the set of simple reflections $\{r_{\alpha_1},\dots
r_{\alpha_8}\}$, and I will write these as $\{r_1,\dots r_8\}$.

Now I want to study the properties of one particular element $w\in W$.
It is a Coxeter element, meaning that it is a product in some order of
a set of simple reflections. Although the ordering is not crucial, the
result I'm after is most transparent if I pick
\[
w=r_3r_4r_6r_7r_1r_2r_5r_8~.
\]
This is a Steinberg ordering: reflections of one colour act
first, followed by those of the other. The ordering amongst the
reflections of like colour is immaterial -- 
they all commute, since the
corresponding simple roots are orthogonal.
The project is to see how $w^{-1}$ acts on $\Phi$, and as a start we can
examine the orbit of $\alpha_1$ under $w^{-1}$. Noting that
\[
\quad r_i(\alpha_j)=\alpha_j-C^{[E_8]}_{ji}\alpha_i~,\qquad
\mbox{(no sum on $j$)}
\]
the individual simple reflection $r_i$ negates $\alpha_i$, adds $\alpha_i$
to all roots $\alpha_j$ joined to $\alpha_i$ by a line on the Dynkin
diagram, and leaves the others alone. With this information it doesn't
take too long to compute that
\[
w^{-1}(\alpha_1)=r_8r_5r_2r_1r_7r_6r_4r_3(\alpha_1)=\alpha_3+\alpha_5~.
\]
\bexercise{ check this!}

\noindent
To continue is easy if a little tedious, acting repeatedly
with $w^{-1}$ to find $w^{-2}(\alpha_1)$, $w^{-3}(\alpha_1)$, and so
on. After $30$ steps, you should find yourself back at $\alpha_1$ (the
number $30$ might just be familiar from the list at the end of the
last lecture). The story for the first $14$ of these steps is
contained in the following table:

\medskip
\[
\baselineskip=9pt
\vbox{
\row{14}{\Y\N\Y\Ns5}
\row{13}{\Ns4\Y\N\Y\N}
\row{12}{\Ns3\Y\N\Y\N\Y}
\row{11}{\N\Y\Ns3\Ys3}
\row{10}{\Ns2\Ys3\N\Ys2}
\row{9}{\Y\N\Y\N\Ys4}
\row{8}{\N\Y\N\Ys5}
\row{7}{\Ns3\Y\N\Ys2\YY}
\row{6}{\N\Ys2\N\Ys4}
\row{5}{\Y\N\Ys3\N\Ys2}
\row{4}{\Ns4\Ys4}
\row{3}{\N\Y\N\Y\N\Y\N\Y}
\row{2}{\Ns6\Ys2}
\row{1}{\Ns2\Y\N\Y\Ns3}
\row{0}{\Y\Ns7}
\bigskip
\hbox{\qquad\qquad\qquad\qquad\quad
 Images of $\alpha_1\!$ under $w^{-1}\,$}
}
\]

\noindent
The coefficient of $\alpha_i$ in the expansion of
$w^{-p}(\alpha_1)$ is given by the
number of blobs ($\bullet$) in the $i^{\rm th}$ position of the
$p^{\rm th}$ row. For the $E_8$ Weyl group, $w^{15}=-1$ and so the
rest of the table, rows $15$ to $29$, can be omitted.

All of this might seem a long way from exact S-matrices,
but in fact the Weyl group computation just performed and the
earlier bootstrap manipulations are in some senses one and
the same calculation, just looked at from orthogonal directions. 
To explain this somewhat delphic remark, I will first rewrite the
S-matrix elements already seen in a new and slightly more compact
notation. Observe that, apart from the blocks $(2)$ and $(28)$ in the
formula for $S_{11}$, every block $(x)$ in $S_{11}$, $S_{12}$ and
$S_{13}$ can be paired off with either $(x{-}2)$ or $(x{+}2)$.
Noticing that $(0)=1$ and $(30)=-1$, this pairing can be extended
to the recalcitrant $S_{11}$ as well, and in fact works for all of
the other S-matrix elements too. Thus we can at least save some ink if
we define a larger building block 
\[
\{x\}=(x{-}1)(x{+}1)~.
\]
and rewrite the S-matrix elements found previously as
\bea
\openup 2\jot
S_{11}&=&\{1\}\{11\}\{19\}\{29\} \nn\\
S_{12}&=&\{7\}\{13\}\{17\}\{23\} \nn\\
S_{13}&=&\{2\}\{10\}\{12\}\{18\}\{20\}\{28\} \nn
\eea
The next step is to introduce a pictorial representation of these
formulae.
Start by drawing a line segment to represent the  interval from
$0$ to $i\pi$ on which the physical-strip poles are found. Then for
each block $\{x\}$ in the S-matrix element, place a small brick 
$\vbox{\hbox to 12pt{\hfil\P\hfil}\hrule}$ on the line segment,
running from $i(x{-}1)\pi/30$ to $i(x{+}1)\pi/30$. (Thus, the poles
are located at the ends of the bricks.) The formulae just given become
\bea
\openup 3\jot
S_{11}~~~&=&~~~\bag{\P\q4\P\q3\P\q4\P} \nn\\
S_{12}~~~&=&~~~\bag{\q3\P\q2\P\q1\P\q2\P\q3} \nn\\
S_{13}~~~&=&~~~\bagh{\P\q3\P\P\q2\P\P\q3\P} \nn
\eea
Rotate these three by $90$ degrees and you should observe a neat match
with the first three columns of the table on the last page,
of images of $\alpha_1$ under $w^{-1}$. 

This is a glimpse of a general construction, which allows a
diagonal scattering theory to be associated with
every simply-laced Weyl group. Further details can be found in Dorey
(1991,1992a); see also Fring and Olive (1992) and Dorey (1992b). 
All of these scattering
theories were in fact already around in the literature: in addition to
the articles by Zamolodchikov already cited, some relevant references
are K\"oberle and Swieca (1979), Sotkov and Zhu (1989),
Fateev and Zamolodchikov (1990),
Christe and Mussardo (1990a,b), Braden et al.~(1990), and Klassen and Melzer
(1990). Why then worry about Weyl groups? This is ultimately a matter of
taste, but it should be mentioned that the construction goes rather
deeper than the curious coincidences described so far. The geometry of
finite reflection groups
appears to replace the rather more complicated Lie algebraic
concepts that might have been a first guess as to the
underlying mathematical structure. Features such as the coupling data
and the pole structure can be related to simple properties of 
root systems, and this allows 
the bootstrap equations both for the conserved currents and for the
S-matrices to be proved in a uniform way.

\smallskip

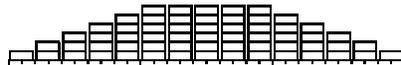
\[
\bag{\P\p2\p3\p4\p5\p6\p6\p6\p6\p6\p5\p4\p3\p2\P}
\]

\section{Affine Toda field theory}
Zamolodchikov's $E_8$-related S-matrix is an example of a diagonal
S-matrix with few of the subtleties that made the treatment of the
sine-Gordon model so delicate. 
Whilst higher poles are certainly present, their
orders are always just as would be predicted from an initial glance at
the possible Landau diagrams. In particular, simple poles are always
associated with bound states. Since the cancellations which
complicated the sine-Gordon case relied on the non-diagonal nature
of its S-matrix, one might suppose that diagonal scattering theories
would always behave in a straightforward manner. Curiously
enough, this turns out not to be true. The affine Toda field theories,
the subject of this lecture, provide a number of elegant
counterexamples.

The study of these models
begins with a standard, albeit non-polynomial, scalar Lagrangian in
1+1 dimensions:
\[
{\cal L}=\frac{1}{2}(\partial\phi)^2-\frac{m^2}{\beta^2}
\sum_{a=0}^rn_ae^{\beta\alpha_a\cdot\phi}.
\]
This describes the interaction of $r$ scalar fields, gathered together
into the vector $\phi\in\bbbr^r$. The set
$\{\alpha_0\dots\alpha_r\}$ is a collection of $r{+}1$ further vectors
in $\bbbr^r$, which must be carefully picked if the model is to be
integrable. It turns out that there is a classically
acceptable choice for every
(untwisted or twisted) affine Dynkin diagram $g^{(k)}$, thought of as
encoding the mutual inner products of the $\alpha_a$. By convention
$\alpha_0$ corresponds to the `extra' spot on the affine diagram, and
the integers $n_a$ satisfy $n_0=1$ and $\sum_{a=0}^rn_a\alpha_a=0$. 
The real constant $m$ sets a mass scale, while $\beta$ governs the
strength of the interactions. When $\beta$ is also
real, the models generalise sinh-Gordon rather
than sine-Gordon and there is no topology to worry about. In fact
once we go beyond the sinh-Gordon example
making $\beta$ imaginary is no longer an innocent operation, since
in all other cases the manifest reality of the
Lagrangian is promptly lost. Despite these problems the models with
$\beta$ purely imaginary have received a fair amount of attention,
starting with the work of Hollowood (1992). However, in
this lecture I will stick to the cases where $\beta$ is real.

As classical field theories, these models are all
integrable, and exhibit conserved quantities at spins given by the
exponents of $g^{(k)}$, repeated modulo a quantity called the $k^{\rm
th}$ Coxeter number, $h^{(k)}$:
\[
h^{(k)}=k\sum_{a=0}^kn_a~.
\]
(For the untwisted diagrams, $k{=}1$ and $h^{(k)}$ is the same as the
usual Coxeter number $h$.) 

However when we turn to the quantum theory, none of the elegant
classical apparatus, as described in, for example, Mikhailov et al.~(1981),
Wilson (1981), and Olive and Turok (1985), is immediately applicable.
A more elementary approach is appropriate, studying the models with
the standard perturbative tools of quantum field theory before
proceeding to some exact conjectures. Arinshtein et al.\
were the first to try this, for the $a_n^{(1)}$ theories, in 1979.
Interest in the subject was renewed following Zamolodchikov's work 
on perturbed conformal field theories, and the fact that in the
meantime the other classically-integrable possibilities, related to
the other affine Dynkin diagrams, had been uncovered. Initially,
only the so-called self-dual models were
understood, and elements of this story can be found in 
Christe and Mussardo (1990a,b) and Braden et al.~(1990,1991).
The other, non self-dual, 
cases were more tricky, since they turned out to fall into
the class of less straightforward
scattering theories for which simple poles do not always
have simple explanations. The crucial step was made by Delius
et al.~(1992), and the papers by Corrigan et al.~(1993) and Dorey (1993)
can be consulted for the few cases not covered in
their work. In the remainder of this lecture I will outline some
aspects of these quantum considerations, but the discussion will
perforce be very sketchy. In addition to the references just cited,
the review by Corrigan (1994) is a good place to start for those
interested in delving deeper into this subject. That the field is
still developing is evinced by an
article by Oota (1997) which appeared as these notes were being
written up, indicating that
the ideas discussed in the last lecture may also be
relevant, if suitably $q$-deformed, to the 
non self-dual theories that had previously resisted any geometrical
interpretation.

If we are to treat these models as ordinary quantum field theories,
then the first step must be to find out what the multipoint couplings
are. To this end, the potential term in the Lagrangian can be expanded
as follows:
\bea
V(\phi)&\,\equiv\,&
\frac{m^2}{\beta^2}\sum_{a=0}^rn_ae^{\beta\alpha_a\cdot\phi}\nn\\
&\,=\,&\frac{m^2}{\beta^2}\sum_{a=0}^rn_a +
\frac{1}{2}(M^2)^{ij}\phi^i\phi^j+
\frac{1}{3!}C^{ijk}\phi^i\phi^j\phi^k +\dots\nn
\eea
where summations on the repeated indices $i$, $j$ and $k$ running from
$1$ to $r$ are implied, and the two and three index objects
\[
(M^2)^{ij}=m^2\sum_{a=0}^rn_a\alpha^i_a\alpha^j_a
\]
and
\[
C^{ijk}=m^2\beta\sum_{a=0}^rn_a\alpha_a^i \alpha_a^j \alpha_a^k
\]
can be thought of as the mass$^2$ matrix and the set of three-point
couplings, at least classically. Much as for theories discussed in the
first lecture, it is possible to view
the $C^{ijk}$ as containing the `bones' of 
the model, with the higher couplings, hidden as `${}+{}\dots\,$',
there just to tidy away any residual production
amplitude backgrounds that would otherwise spoil integrability.
Now the general idea is the
following: first diagonalise $M^2$ to find the classical particle
masses $m_1\dots m_r$, and then compute the $C^{ijk}$ in the
eigenbasis of $M^2$ to find the classical three-point couplings 
between the corresponding one-particle states. 
At this level there are already some surprises: for
example, it turns out that in all of the untwisted cases, the set of
masses form the eigenvector, with lowest eigenvalue, of the
corresponding non-affine Cartan matrix. This was initially
noticed on a case-by-case basis, before being proved in a general way
by Freeman (1991). The Coxeter element, described in the last lecture,
turns out to be crucial in this discussion.
This work was further elaborated by Fring et al.~(1991), elucidating
in particular earlier observations about the three-point couplings.

However it has been obtained, once the classical data is known
two things can be done:
on the one hand the masses and three-point couplings can be fed into
the bootstrap to make some initial conjectures as to the
full quantum S-matrices, and on the other 
perturbation theory can be attempted in order to check these
conjectures. As hinted above, the affine Toda field theories split
into two classes when this programme is attempted: `straightforward'
and `not straightforward'. To make this distinction more precise,
define a duality operation on the set of all affine Dynkin diagrams
by
\[
\{\alpha_0\dots\alpha_r\}\leftrightarrow
\{\alpha_0^{\vee}\dots\alpha_r^{\vee}\}
\]
where
\[
\alpha_a^{\vee}\equiv\frac{2}{\alpha_a^2}\alpha_a\,.
\]
(This is sometimes called Langlands duality.)

When appropriately normalised, the sets of vectors associated with the
$a^{(1)}_n$, $d^{(1)}_n$, $e^{(1)}_n$ and $a^{(2)}_{2n}$ affine Dynkin
diagrams are self-dual in this sense, and the corresponding affine
Toda field theories are also called self-dual. These are the
`straightforward' cases: conjectures based on the classical data lead
to self-consistent quantum S-matrices, and to date these have passed all
perturbative checks to which they have been subjected. For example,
the mass ratios, and hence the fusing angles, are preserved at one
loop. As a result, the bootstrap structure is essentially blind to the
value of the coupling $\beta$, which enters into the S-matrices 
via a function
\[
B(\beta)=\frac{1}{2\pi}\frac{\beta^2}{1+\beta^2/4\pi}~.
\]
There is a simple relationship between the S-matrices for 
certain perturbed conformal field theories and
the S-matrices for the
self-dual affine Toda models: all that has to be done is to
replace building blocks of the type seen in the last lecture
\[
\{x\}_{\rm PCFT}\equiv (x{-}1)(x{+}1)
\]
by the slightly more elaborate blocks
\[
\{x\}_{\rm toda}\equiv\frac{(x{-}1)(x{+}1)}{(x{-}1{+}B)(x{+}1{-}B)}\,.
\]
Zamolodchikov's $E_8$-related S-matrix is 
related in this way to the S-matrix
of the $e^{(1)}_8$ affine Toda field theory; more generally, for
$g\in\{a,d,e\}$ the correspondence is between the $g$ affine Toda
field theory and a perturbation of the $g^1{\times}g^1/g^2$ coset
model,
while $a^{(2)}_{2n}$ turns into a perturbation of the
nonunitary minimal model ${\cal M}(2,2n{+}3)$.
(Note though that the factors of $60$ appearing in the
earlier definition of $(x)$ should be replaced by $2h$, with $h$ the
relevant Coxeter number). For every self-dual affine Toda
S-matrix, there is thus a companion `minimal' S-matrix, sharing the 
same physical pole structure but lacking the
coupling-constant dependent physical strip zeroes, which in the
Toda theories serve to cancel the poles in the 
$\beta\rightarrow 0$ limit.
Note
also that replacing $\beta$ by $4\pi/\beta$ sends $B$ to $2{-}B$ and
leaves the Toda blocks unchanged -- a strong-weak coupling duality.

The remaining, non self-dual, models behave in a much more complicated
way. The classical data is still elegant, but there is quantum
trouble: conjectures based on the raw classical data are no longer
self-consistent, and perturbative checks show varying mass ratios,
causing the fusing angles to depend on the value
of the coupling constant. These perturbative results are reinforced
by the results of Kausch and Watts (1992) and Feigin and Frenkel (1993),
which indicate that the correct general implementation of strong-weak
coupling duality is not only to replace $\beta$ by $4\pi/\beta$, but
also to replace each $\alpha_a$ by its dual, $\alpha_a^{\vee}$. (Of
course, for the self-dual theories this latter operation has no effect
and so the earlier statement of duality remains correct for these
cases.) This means that for each dual pair of classical affine Toda
field theories, there should be just one quantum theory -- there are
`fewer' genuinely distinct quantum theories than expected, 
and the different
classical theories can be recovered by taking strong or weak coupling
limits. The predicted dualities are:
\bea
&&b^{(1)}_n\leftrightarrow a^{(2)}_{2n-1}\qquad\qquad
g^{(1)}_2\leftrightarrow d^{(3)}_4 \nn\\
&&c^{(1)}_n\leftrightarrow d^{(2)}_{n+1}\qquad\qquad~
f^{(1)}_4\leftrightarrow e^{(2)}_6 \nn
\eea
\bexercise{ compare and contrast the classical conserved charge
fingerprints for these models. Are they compatible with duality?}

If this picture is correct, then the mass ratios have no
option but to vary: if a model is non self-dual, then (as can be
checked case-by-case) its classical
mass spectrum is always different from that of its dual. One spectrum
is found at $\beta\rightarrow 0$, the other at
$\beta\rightarrow\infty$, and the quantum theory, if it
exists at all, must find some way of interpolating between the two.
Given the apparent rigidity of the bootstrap equations, this looks to
be rather a tall order. Nevertheless, Delius et al.~(1992) decided to
take the perturbatively-calculated shifts in the mass ratios seriously, 
and were led to a set of conjectures for most of the models in the
above list (the only cases that remained were $d^{(3)}_4$, $e^{(2)}_6$
and $f^{(1)}_4$, and these were subsequently found to behave in just
the same way). Whenever conjectures existed for both halves of a dual
pair, they swapped over under $\beta\rightarrow 4\pi/\beta$. In fact,
Delius et al.\ made their proposals independently of any expectations
of duality; that it emerged anyway from their calculations can be seen
in retrospect as strong evidence that they were on the right track.
Further support came from the numerical results of Watts and Weston
(1992), who examined the coupling-dependence of the single mass
ratio found in the $g^{(1)}_2/d^{(3)}_4$ dual pair. 

With the mass ratios depending on the coupling, it is no longer
possible to use the simple building blocks $\{x\}_{\rm toda}$
introduced earlier. A slightly more elaborate two-index block
$\{x,y\}$ can be found in Dorey (1993), and is probably the most direct
generalisation of the self-dual blocks. One point to note is that the
natural way for the coupling to enter these blocks requires the
previous definition of the function $B(\beta)$ to be slightly
modified, so as to encompass the non self-dual theories as well:
\[
B(\beta)=\frac{1}{2\pi}\frac{\beta^2}{h/h^{\vee}+\beta^2/4\pi}~.
\]
Here $h$ is the Coxeter number of the relevant affine Dynkin diagram,
and $h^{\vee}$ that of its (Langlands) dual.

There are, however, a couple of features of the non self-dual
S-matrices which give pause for thought. Some simple poles, expected
on the basis of the nonzero classical couplings and quantum mass
ratios, turn out to be absent, whilst other simple poles, which are present
in the quantum S-matrices, are not at locations which match any of the
particle masses. 

The resolution of the first problem appears to be that
quantum corrections exactly cancel some of the classical three-point
couplings, when evaluated on shell. This means that some quantum
couplings $C^{ijk}$ vanish even though their classical counterparts do
not.  The result is very delicate and has only
been checked to one loop, but is probably necessary if duality is to
hold, for the simple reason that the set of classical three-point
couplings in a theory and its dual do not in general coincide. Those 
couplings which show signs of 
vanishing once quantum effects are taken into account are
precisely those which are anyway absent at the classical level in the
dual model. 

As for the second problem, the mechanism is not too far
removed from that operating in the sine-Gordon model. However, as
mentioned at the beginning of this lecture, the fact that the
scattering is diagonal means that we can no longer hope
to generate simple poles through cancellations between competing Landau
diagrams. Fortunately there is a compensating feature of the affine
Toda S-matrices which allows the basic idea to be saved: 
they all exhibit zeroes as well as poles on the physical strip. These
were already visible in the self-dual blocks $\{x\}_{\rm toda}$ defined
earlier, and are equally present in the more general blocks $\{x,y\}$. In
the self-dual cases the zeroes are merely spectators, but in
the non self-dual theories they come to play a much more central
role in the pole analysis. A detailed discussion can be found in
Corrigan et al.~(1993), and it turns out that for every `anomalous'
simple pole in the non self-dual affine Toda S-matrices, Landau
diagrams can be drawn in which some internal lines cross.
Just as for sine-Gordon, the S-matrix elements for these internal
crossings must be
factored into the calculation before the overall order of any pole can
be predicted. This time, these factors vanish individually as the
diagrams are put on shell, and
thereby manage to demote ostensibly higher poles into the simple poles
that are required in order to match the quantum S-matrices. 

\section*{Further reading, and acknowledgements}
As promised, these lectures have only skimmed the surface of a large
subject. In addition to the references mentioned in the main text, the 
review articles by Zamolodchikov and Zamolodchikov (1979),
Zamolodchikov (1980) and Mussardo (1992) are recommended, 
as is the discussion of
the sine-Gordon model given by Goebel (1986). (The opening section
the first lecture was in fact inspired by a remark in this article.)

\smallskip\noindent
I would like to thank Zal\'an Horv\'ath and Laci Palla for
all their efforts in organising the 1996 E\"otv\"os Graduate School,
where this material was first presented. These notes are a
slightly expanded (and corrected) version of my contribution to the
proceedings of that school; needless to say, I would be grateful to
hear of any errors and/or typos that remain. Thanks also go to
Olivier Babelon, Jean-Bernard Zuber and the other organisers of the
the `Integrable systems' semester held
at the Institut Henri Poincar\'e, Paris, for giving me a second
opportunity to talk about exact S-matrices during November and
December of 1996.
I am grateful to Jacques Bross, Ed Corrigan, Carlos Fernandez-Pousa,
Daniel Iagolnitzer, G\'erard Watts and especially
Jean-Bernard Zuber for
helpful comments and discussions, and 
to the UK EPSRC for an advanced fellowship.
This work was supported in part by a TMR grant of the European Union,
contract reference  ERBFMRXCT960012.

%%%%%%%%%%%%%%%%%%%%%%%%%%%%%%%%%%%%%%%%%%%%%%%%%%%%
%
% ---- Bibliography ----
%


\begin{thebibliography}
%
\bibitem{}{AKa}{}
Aref'eva,\,I.Ya.~and Korepin,\,V.E.~(1974):
Scattering in two-dimensional model with Lagrangian 
$L=(1/\gamma)\left[(1/2)(\partial_{\mu}u)^2+m^2(\cos u-1)\right]$,
\JETPL{20}, 312--314
%
\bibitem{}{AFZa}{}
Arinshtein,\,A.E., Fateyev,\,V.A.~and Zamolodchikov,\,A.B.~(1979):
Quantum S-matrix of the (1+1)-dimensional Todd chain,
\PL{B87}, 389--392
%
\bibitem{}{BLa}{}
Bernard,\,D.~and Leclair,\,A.~(1991):
Quantum group symmetries and nonlocal currents in 2-D QFT,
\CMP{142}, 99--138
%
\bibitem{}{BCDSa}{}
Braden,\,H.W., Corrigan.\,E., Dorey,\,P.E.~and Sasaki,\,R.~(1990):
Affine Toda field theory and exact S-matrices,
\NP{B338}, 689--746
%
\bibitem{}{BCDSb}{}
Braden,\,H.W., Corrigan.\,E., Dorey,\,P.E.~and Sasaki,\,R.~(1991):
Multiple poles and other features of affine Toda field theory,
\NP{B356}, 469--498
%
\bibitem{}{Caa}{}
Chandler,\,C.~(1969):
Causality in $S$-matrix theory, II,
\HPA{42}, 759--765
%
\bibitem{}{CMb}{}
Christe,\,P.~and Mussardo,\,G.~(1990a):
Integrable systems away from criticality: the Toda field theory and
S-matrix of the tricritical Ising model,
\NP{B330}, 465--487
%
\bibitem{}{CMc}{}
Christe,\,P.~and Mussardo,\,G.~(1990b):
Elastic S-matrices in (1+1) dimensions and Toda field theories,
\IJMP{A5}, 4581--4627
%
\bibitem{}{CMa}{}
Coleman,\,S.~and Mandula,\,J.~(1967):
All possible symmetries of the S-matrix,
\PR{159}, 1251--1256
%
\bibitem{}{CTa}{}
Coleman,\,S.~and Thun,\,H.J.~(1978):
On the prosaic origin of the double poles in the sine-Gordon S-matrix,
\CMP{61}, 31--39
%
\bibitem{}{Ca}{}
Corrigan,\,E.~(1994):
Recent developments in affine Toda quantum field theory,
Invited lectures at the CRM-CAP Summer School {\it Particles and Fields
94}, Banff, Alberta, Canada; hep-th/9412213
%
\bibitem{}{CDSa}{}
Corrigan,\,E., Dorey,\,P.E.~and Sasaki,\,R.~(1993):
On a generalised bootstrap principle,
\NP{B408}, 579--599
%
\bibitem{}{DHNa}{}
Dashen,\,R.F., Hasslacher,\,B.~and Neveu,\,A.~(1975):
The particle spectrum in model field theories from semiclassical
functional integral techniques,
\PR{D11}, 3424--3450
%
\bibitem{}{DGZa}{}
Delius,\,G.W., Grisaru,\,M.T.~and Zanon,\,D.~(1992):
Exact S-matrices for nonsimply-laced affine Toda theories,
\NP{B382}, 365--406
%
\bibitem{}{Da}{}
Dorey,\,P.~(1991):
Root systems and purely elastic S-matrices,
\NP{B358}, 654--676
%
\bibitem{}{Db}{}
Dorey,\,P.E.~(1992a):
Root systems and purely elastic S-matrices (II),
\NP{B374}, 741--761
%
\bibitem{}{Dbb}{}
Dorey,\,P.~(1992b):
Hidden geometrical structures in integrable models,
in the proceedings of the NATO ARW {\it Integrable Quantum Field
Theories}, Como, Italy (Plenum 1993); hepth/9212143
%
\bibitem{}{Dc}{}
Dorey,\,P.~(1993):
A remark on the coupling dependence in affine Toda field theories,
\PL{B312}, 291--298
%
\bibitem{}{FZa}{}
Fateev,\,V.A.~and Zamolodchikov,\,A.B.~(1990):
Conformal field theory and purely elastic S-matrices,
\IJMP{A5}, 1025--1048
%
\bibitem{}{FFa}{}
Feigin,\,B.L.~and Frenkel,\,E.V.~(1993):
Integrals of motion and quantum groups,
Lectures given at the CIME Summer School {\it Integrable Systems and
Quantum Groups}, Montecatini Terme, Italy; hepth/9310022
%
\bibitem{}{Fa}{}
Freeman,\,M.D.~(1991):
On the mass spectrum of affine Toda field theory,
\PL{B261}, 57--61
%
\bibitem{}{FLOa}{}
Fring,\,A., Liao,\,H.C.~and Olive,\,D.I.~(1991):
The mass spectrum and coupling in affine Toda theories,
\PL{B266}, 82--86
%
\bibitem{}{FOa}{}
Fring,\,A.~and Olive,\,D.I.~(1992):
The fusing rule and the scattering matrix of affine Toda theory,
\NP{B379}, 429--447
%
\bibitem{}{Ga}{}
Goebel,\,C.J.~(1986):
On the sine-Gordon S-matrix,
\PTPS{86}, 261--273
%
\bibitem{}{Ha}{}
Hollowood,\,T.~(1992):
Solitons in affine Toda field theories,
\NP{B384}, 523--540
%
\bibitem{}{Iaa}{}
Iagolnitzer,\,D.~(1973):
{\it Introduction to S-matrix theory} (ADT, Paris 1973)
%
\bibitem{}{Ia}{}
Iagolnitzer,\,D.~(1978a):
{\it The S matrix} (North-Holland 1978)
%
\bibitem{}{Ib}{}
Iagolnitzer,\,D.~(1978b):
Factorization of the multiparticle $S$ matrix in two-dimensional
space-time models,
\PR{D18}, 1275-1285
%
\bibitem{}{KWa}{}
Kausch,\,H.G.~and Watts,\,G.M.T.~(1992):
Duality in quantum Toda theory and W algebras,
\NP{B386}, 166-192
%
\bibitem{}{KMa}{}
Klassen,\,T.R.~and Melzer,\,E.~(1990):
Purely elastic scattering theories and their ultraviolet limits,
\NP{B338}, 485--528
%
\bibitem{}{KSa}{}
K\"oberle,\,R.~and Swieca,\,J.A.~(1979):
Factorizable $Z(N)$ models,
\PL{B86}, 209--210
%
\bibitem{}{KFa}{}
Korepin,\,V.E.~and Faddeev,\,L.D.~(1975):
Quantization of solitons,
\TMF{25}, 147--163
%
\bibitem{}{La}{}
L\"uscher,\,M.~(1978):
Quantum non-local charges and absence of particle production in the
two-dimensional non-linear $\sigma$-model,
\NP{B135}, 1--19
%
\bibitem{}{MWa}{}
McCoy,\,B.M.~and Wu,\,T.T.~(1978):
Two-dimensional Ising field theory in a magnetic field: Breakup of the
cut in the two-point function,
\PR{D18}, 1259--1267
%
\bibitem{}{MOPa}{}
Mikhailov,\,A.V., Olshanetsky,\,M.A.~and Perelomov,\,A.M.~(1981):
Two-dimensional generalised Toda lattice,
\CMP{79}, 473--488
%
\bibitem{}{Ma}{}
Mussardo,\,G.~(1992):
Off-critical statistical models: factorized scattering theories and
bootstrap program,
\PRP{218}, 215--379
%
\bibitem{}{OTa}{}
Olive,\,D.I.~and Turok,\,N.~(1985):
Local conserved densities and zero-curvature conditions for Toda
lattice field theories,
\NP{B257}, 277--301
%
\bibitem{}{Oa}{}
Oota,\,T.~(1997):
$q$-deformed Coxeter element in non-simply-laced affine Toda field
theories,
Yukawa Institute preprint YITP-97-33, hepth/9706054
%
\bibitem{}{Pb}{}
Parke,\,S.~(1980):
Absence of particle production and factorization of the S matrix in
(1+1)-dimensional models,
\NP{B174}, 166--182
%
\bibitem{}{Pa}{}
Polyakov,\,A.M.~(1977):
Hidden symmetry of the two-dimensional chiral fields,
\PL{B72}, 224--226
%
\bibitem{}{SWa}{}
Shankar,\,R.~and Witten,\,E.~(1978):
$S$ matrix of the supersymmetric nonlinear $\sigma$ model,
\PR{D17}, 2134--2143
%
\bibitem{}{SZa}{}
Sotkov,\,G.~and Zhu,\,C.-J.~(1989):
Bootstrap fusions and tricritical Potts model away from criticality,
\PL{B229}, 391--397
%
\bibitem{}{Ta}{}
Tzitz\'eica,\,M.~(1910):
Sur une nouvelle classe de surfaces,
Comptes Rendus Acad.\ Sci.\ {\bf 150}, 955
%
\bibitem{}{WWa}{}
Watts,\,G.M.T.~and Weston,\,R.A.~(1992):
$G^{(1)}_2$ affine Toda field theory. A numerical test of exact
S-matrix results,
\PL{B289}, 61--66
%
\bibitem{}{Wa}{}
Wilson,\,G.~(1981):
The modified Lax and two-dimensional Toda lattice equations associated
with simple Lie algebras,
\ETDS{1}, 361--380
%
\bibitem{}{Za}{}
Zamolodchikov,\,A.B.~(1977):
Exact two-particle S-matrix of quantum sine-Gordon solitons,
\CMP{55}, 183--186
%
\bibitem{}{Zaa}{}
Zamolodchikov,\,A.B.~(1980):
Factorised S matrices and lattice statistical systems,
Sov.~Sci.~Rev., Physics, {\bf 2}, 1--40
%
\bibitem{}{Zb}{}
Zamolodchikov,\,A.B.~(1989a):
Integrals of motion and S-matrix of the (scaled) $T{=}T_c$ Ising model
with magnetic field,
\IJMP{A4}, 4235--4248
%
\bibitem{}{Zc}{}
Zamolodchikov,\,A.B.~(1989b):
Integrable field theory from conformal field theory,
Advanced Studies in Pure Mathematics {\bf 19}, 641--674
%
\bibitem{}{Zd}{}
Zamolodchikov,\,A.B.~(1989c):
Fractional-spin integrals of motion in perturbed conformal field
theory,
in {\it Fields, strings and quantum gravity}, proceedings of the CCAST
symposium workshop, Beijing 1989 (Gordon and Breach)
%
\bibitem{}{Ze}{}
Zamolodchikov,\,Al.B.~(1990):
Thermodynamic Bethe ansatz in relativistic models. Scaling 3-state
Potts and Lee-Yang models,
\NP{B342}, 695--720
%
\bibitem{}{ZZa}{}
Zamolodchikov,\,A.B.~and Zamolodchikov,\,Al.B.~(1979):
Factorised S-matrices in two dimensions as the exact solutions of
certain relativistic quantum field theory models,
\AP{120}, 253--291
%
\end{thebibliography}
\end{document}